\documentclass[12pt]{article}
\usepackage{epsfig}
\begin{document}
\title{Towards clarifying the possibility of observation of the LHCb hidden-charm pentaquarks\\
$P_{c}^+(4312)$, $P_{c}^+(4337)$, $P_{c}^+(4440)$ and $P_{c}^+(4457)$\\ in near-threshold charmonium
photoproduction off protons and nuclei}
\author{E. Ya. Paryev \\
{\it Institute for Nuclear Research, Russian Academy of Sciences,}\\
{\it Moscow 117312, Russia}}

\renewcommand{\today}{}
\maketitle

\begin{abstract}
We study the near-threshold $J/\psi$ meson photoproduction from protons and nuclei
by considering incoherent direct non-resonant
(${\gamma}p \to {J/\psi}p$, ${\gamma}n \to {J/\psi}n$) and two-step resonant
(${\gamma}p \to P_{ci}^+ \to {J/\psi}p$, ${\gamma}n \to P_{ci}^0 \to {J/\psi}n$, $i=1$, 2, 3, 4;
$P_{c1}^{+,0}=P_c^{+,0}(4312)$, $P_{c2}^{+,0}=P_c^{+,0}(4337)$, $P_{c3}^{+,0}=P_c^{+,0}(4440)$,
$P_{c4}^{+,0}=P_c^{+,0}(4457)$) charmonium production processes.
We calculate the absolute excitation functions, energy and momentum distributions
for the non-resonant, resonant and for the combined (non-resonant plus resonant) production
of $J/\psi$ mesons on protons as well as, using the nuclear spectral function approach,
on carbon and tungsten target nuclei at near-threshold incident photon energies by assuming
the spin-parity assignments of the hidden-charm resonances $P_{c}^{+,0}(4312)$,
$P_{c}^{+,0}(4337)$, $P_{c}^{+,0}(4440)$ and $P_{c}^{+,0}(4457)$ as $J^P=(1/2)^-$, $J^P=(1/2)^-$, $J^P=(1/2)^-$
and $J^P=(3/2)^-$ within three different realistic scenarios for the branching ratios
of their decays to the ${J/\psi}p$ and ${J/\psi}n$ modes (0.25, 0.5 and 1\%).
We show that will be very hard to measure the $P_{ci}^+$ pentaquark states through the scan of the $J/\psi$
total photoproduction cross section on a proton target in the near-threshold energy region around the resonant
photon energies of 9.44, 9.554, 10.04 and 10.12 GeV if these branching ratios $\sim$ 1\% and less.
We also demonstrate that at these photon beam energies the $J/\psi$ energy and momentum combined distributions
considered reveal distinct sensitivity to the above scenarios, respectively, at "low" $J/\psi$ total energies and momenta, which implies that they may be an important tool to provide further evidence for the existence of
the pentaquark $P_{ci}^+$ and $P_{ci}^0$ resonances and to get valuable information on their decay rates to the ${J/\psi}p$ and ${J/\psi}n$ final states. The measurements of these distributions could be performed in the future
at the 12 GeV CEBAF facility.
\end{abstract}

\newpage

\section*{1. Introduction}
The study of the exotic hadronic states -- the hidden-charm pentaquark resonances -- has received considerable
interest in recent years and is one of the most exciting topics of the nuclear and hadronic physics nowadays
after the discovery by the LHCb Collaboration pentaquark states $P_c^+(4380)$ and $P_c^+(4450)$ in the ${J/\psi}p$  invariant mass spectrum of the $\Lambda^0_b \to K^-({J/\psi}p)$ decays [1] and, especially, after the observation
by the Collaboration of three new narrow resonances $P_c^+(4312)$, $P_c^+(4440)$ and $P_c^+(4457)$ in these
decays [2], based on additional collected data and on an improved selection strategy, instead of initially
claimed $P_c^+(4380)$ and $P_c^+(4450)$ states.
Very recently, the LHCb Collaboration discovered a new narrow hidden-charm pentaquark denoted as
$P_{c}^+(4337)$ in the invariant mass spectrum of the ${J/\psi}p$ in the $B_s^0 \to {J/\psi}p{\bar p}$ decays [3].
On the other hand, the search for the LHCb pentaquarks $P_c^+(4312)$, $P_c^+(4440)$ and $P_c^+(4457)$ by the GlueX Collaboration in Hall D at JLab through a scan of the cross section of elastic reaction ${\gamma}p \to {J/\psi}p$ from threshold of 8.21 GeV and up to incident photon energy $E_{\gamma}=11.8$ GeV gave no evidence for them with present statistics and set the model-dependent  upper limits on branching ratios of $P_c^+(4312) \to {J/\psi}p$, $P_c^+(4440) \to {J/\psi}p$ and $P_c^+(4457) \to {J/\psi}p$ decays of several percent [4]. The preliminary results from a factor of 10 more data on the $J/\psi$ photoproduction on a hydrogen target,
collected in the Hall C JLab E12-16-007 experiment (the so-called $J/\psi$-007 experiment),
also show no $P_c^+$ signal [5]. In this experiment the $e^+e^-$ pairs from the $J/\psi$ decays were detected
in coincidence using the two high momentum spectrometers of Hall C: the Super High Momentum Spectrometer
(SHMS) and High Momentum Spectrometer (HMS) for the electron and the positron, respectively.
In recent publications [6] and [7] the role, respectively, of the LHCb pentaquarks
$P_c^+(4450)$ and $P_c^+(4312)$, $P_c^+(4440)$, $P_c^+(4457)$ in charmonium photoproduction on protons and nuclei
has been investigated at near-threshold initial photon energies $E_{\gamma} \le 11$ GeV. Here, the description was
based on the consideration of the incoherent direct (${\gamma}N \to {J/\psi}N$) and two-step
(${\gamma}p \to P_c^+(4450) \to {J/\psi}p$ and ${\gamma}p \to P_c^+(4312) \to {J/\psi}p$,
${\gamma}p \to P_c^+(4440) \to {J/\psi}p$, ${\gamma}p \to P_c^+(4457) \to {J/\psi}p$) $J/\psi$ production
processes. As a measure for this role, the incident photon energy dependence of $J/\psi$ production
cross sections on protons and nuclei (excitation functions) has been adopted.
It was found that it is insignificant for the pentaquark resonances considered if branching ratios of their
decays to the ${J/\psi}p$ mode are less than a few percent, which is in line with the results of the JLab
experiments [4, 5].

 In view of the aforementioned, to get a robust enough information for or against the existence of the
LHCb hidden-charm pentaquarks and to understand their better, it is crucial to investigate the possibility
of their observation by measuring not only the excitation functions for $J/\psi$ meson production
from photon-induced reactions on protons and nuclei at near-threshold photon energies, predicted in Refs. [6, 7],
but also the $J/\psi$ energy and momentum distributions in these reactions, not predicted in the previous
papers [6, 7]. Their prediction is the main aim of the present study.
In it, we consider the contribution of the $P_{c}^{+,0}(4312)$, $P_{c}^{+,0}(4337)$,
$P_{c}^{+,0}(4440)$ and $P_{c}^{+,0}(4457)$ resonances to near-threshold $J/\psi$ photoproduction
off protons and nuclei by adopting the Breit-Wigner shape for this contribution and by employing
the recent experimental data [4] on the total and differential cross sections of the ${\gamma}p \to {J/\psi}p$
process to estimate the background contribution.
The consideration is mainly based on the model, developed in Refs. [6, 7].
We present the predictions obtained within our present approach for the $J/\psi$ energy and momentum distributions
in ${\gamma}p$ as well as in ${\gamma}$$^{12}$C and ${\gamma}$$^{184}$W reactions at
near-threshold incident photon energies. These predictions may be useful in planning future
$J/\psi$ photoproduction experiments at the CEBAF facility.

\section*{2. Theoretical framework}

\subsection*{2.1. Direct non-resonant $J/\psi$ production processes}

  At near-threshold photon beam energies below 11 GeV of our interest
\footnote{$^)$These energies are well within the present capabilities  of the upgraded CEBAF facility
at JLab, which is providing an opportunity to study the observed [1--3] by the LHCb Collaboration
exotic hidden-charm pentaquark states in exclusive ${\gamma}p \to {J/\psi}p$ reactions in all
experimental Halls A, B, C, D [4, 5].}$^)$,
the following direct non-resonant elementary charmonium production processes with the lowest free production
threshold ($\approx$~8.21 GeV) contribute to the $J/\psi$ photoproduction on nuclei [6--8]:
\begin{equation}
\gamma+p \to J/\psi+p,
\end{equation}
\begin{equation}
\gamma+n \to J/\psi+n.
\end{equation}
The modification of the masses of the final high-momentum
$J/\psi$ meson and proton (see below) in the nuclear medium will be ignored in the present study.

Disregarding the absorption of incident photons in the energy range of interest as well as the $J/\psi$
meson quasielastic rescatterings on target nucleons [9], and describing
the charmonium final-state absorption in the nuclear matter by the
absorption cross section $\sigma_{{J/\psi}N}$
\footnote{$^)$For which we will use the value $\sigma_{{J/\psi}N}=3.5$ mb [6--10].}$^)$,
we represent the inclusive differential cross section for the production of $J/\psi$ mesons with the
momentum ${\bf p}_{J/\psi}$ on nuclei in the direct non-resonant processes (1), (2)
in the form [6--10]:
\begin{equation}
\frac{d\sigma_{{\gamma}A\to {J/\psi}X}^{({\rm dir})}({\bf p}_{\gamma},{\bf p}_{J/\psi})}
{d{\bf p}_{J/\psi}}=I_{V}[A,\sigma_{{J/\psi}N}]
\left<\frac{d\sigma_{{\gamma}p \to {J/\psi}p}({\bf p}_{\gamma},{\bf p}_{J/\psi})}{d{\bf p}_{J/\psi}}\right>_A,
\end{equation}
where
\begin{equation}
I_{V}[A,\sigma]=2{\pi}A\int\limits_{0}^{R}r_{\bot}dr_{\bot}
\int\limits_{-\sqrt{R^2-r_{\bot}^2}}^{\sqrt{R^2-r_{\bot}^2}}dz
\rho(\sqrt{r_{\bot}^2+z^2})
\exp{\left[-A{\sigma}\int\limits_{z}^{\sqrt{R^2-r_{\bot}^2}}\rho(\sqrt{r_{\bot}^2+x^2})dx\right]},
\end{equation}
\begin{equation}
\left<\frac{d\sigma_{{\gamma}p \to {J/\psi}p}({\bf p}_{\gamma},{\bf p}_{J/\psi})}{d{\bf p}_{J/\psi}}\right>_A=
\int\int
P_A({\bf p}_t,E)d{\bf p}_tdE
\left[\frac{d\sigma_{{\gamma}p \to {J/\psi}p}(\sqrt{s^*},{\bf p}_{J/\psi})}{d{\bf p}_{J/\psi}}\right]
\end{equation}
and
\begin{equation}
  s^*=(E_{\gamma}+E_t)^2-({\bf p}_{\gamma}+{\bf p}_t)^2,
\end{equation}
\begin{equation}
   E_t=M_A-\sqrt{(-{\bf p}_t)^2+(M_{A}-m_{p}+E)^{2}}.
\end{equation}
Here, $d\sigma_{{\gamma}p\to {J/\psi}p}(\sqrt{s^*},{\bf p}_{J/\psi})/d{\bf p}_{J/\psi}$
is the off-shell differential cross section for the production of $J/\psi$ meson in process (1)
at the "in-medium" ${\gamma}p$ c.m. energy $\sqrt{s^*}$
\footnote{$^)$In Eq. (3), it is assumed that the cross sections for $J/\psi$ meson production in
${\gamma}p$ and ${\gamma}n$ interactions are equal to each other [6--8].}$^)$;
$\rho({\bf r})$ and $P_A({\bf p}_t,E)$ are normalized to unity the nucleon density and the nuclear
spectral function (they are given in Refs. [11, 12]) of target nucleus with mass number $A$,
having mass $M_{A}$ and radius $R$;
${\bf p}_{\gamma}$ and $E_{\gamma}$ are the momentum and
energy of the incident photon ($E_{\gamma}=|{\bf p}_{\gamma}|=p_{\gamma}$) in the laboratory system;
${\bf p}_{t}$  and $E$ are the momentum and binding energy of the intranuclear target proton,
participating in the reaction channel (1); $m_p$ is the free space proton mass.

 Also, as previously in [6--8], we will suppose that the off-shell differential cross section\\
$d\sigma_{{\gamma}p\to {J/\psi}p}(\sqrt{s^*},{\bf p}_{J/\psi})/d{\bf p}_{J/\psi}$
for $J/\psi$ production in process (1) is the same as the corresponding on-shell cross section
$d\sigma_{{\gamma}p\to {J/\psi}p}(\sqrt{s},{\bf p}_{J/\psi})/d{\bf p}_{J/\psi}$
determined for the off-shell kinematics of this process and in which the vacuum c.m. energy squared $s$,
given by the formula
\begin{equation}
  s=W^2=(E_{\gamma}+m_p)^2-{\bf p}_{\gamma}^2=m_p^2+2m_pE_{\gamma},
\end{equation}
is replaced by the in-medium expression (6). The above off-shell differential cross section is then
(cf. [8, 10]):
\begin{equation}
\frac{d\sigma_{{\gamma}p \to {J/\psi}p}(\sqrt{s^*},{\bf p}_{J/\psi})}
{d{\bf p}_{J/\psi}}=
\frac{\pi}{I_2(s^*,m_{J/\psi},m_{p})E_{J/\psi}}
\end{equation}
$$
\times
\frac{d\sigma_{{\gamma}p \to {J/\psi}{p}}(\sqrt{s^*},\theta_{J/\psi}^*)}{d{\bf \Omega}_{J/\psi}^*}
\frac{1}{(\omega+E_t)}\delta\left[\omega+E_t-\sqrt{m_{p}^2+({\bf Q}+{\bf p}_t)^2}\right],
$$
where
\begin{equation}
I_2(s^*,m_{J/\psi},m_{p})=\frac{\pi}{2}
\frac{\lambda(s^*,m_{J/\psi}^{2},m_{p}^{2})}{s^*},
\end{equation}
\begin{equation}
\lambda(x,y,z)=\sqrt{{\left[x-({\sqrt{y}}+{\sqrt{z}})^2\right]}{\left[x-
({\sqrt{y}}-{\sqrt{z}})^2\right]}},
\end{equation}
\begin{equation}
\omega=E_{\gamma}-E_{J/\psi}, \,\,\,\,{\bf Q}={\bf p}_{\gamma}-{\bf p}_{J/\psi},\,\,\,\,
E_{J/\psi}=\sqrt{m^2_{J/\psi}+{\bf p}_{J/\psi}^2}.
\end{equation}
Here, the off-shell c.m. charmonium angular distribution in reaction (1)
$d\sigma_{{\gamma}p \to {J/\psi}{p}}(\sqrt{s^*},\theta_{J/\psi}^*)/d{\bf \Omega}_{J/\psi}^*$
as a function of the $J/\psi$ production c.m. polar angle $\theta_{J/\psi}^*$ is given by [10]:
\begin{equation}
\frac{d\sigma_{{\gamma}p \to {J/\psi}p}(\sqrt{s^*},\theta^*_{J/\psi})}{d{\bf \Omega}_{J/\psi}^*}=
a{\rm e}^{b_{J/\psi}(t-t^+)}\sigma_{{\gamma}p \to {J/\psi}p}(\sqrt{s^*}),
\end{equation}
where $\sigma_{{\gamma}p \to {J/\psi}p}(\sqrt{s^*})$ is the off-shell total cross section
for $J/\psi$ meson production in this reaction and
\begin{equation}
t=m_{J/\psi}^2-2E^*_{\gamma}E^*_{J/\psi}+2p^*_{\gamma}p^*_{J/\psi}\cos{\theta^*_{J/\psi}},
\end{equation}
\begin{equation}
E_{\gamma}^*=p^{*}_{\gamma}, \,\,\,\,
E_{J/\psi}^*=\sqrt{m^2_{J/\psi}+p^{*2}_{J/\psi}},
\end{equation}
\begin{equation}
p_{\gamma}^*=\frac{1}{2\sqrt{s^*}}\lambda(s^*,0,E_{t}^2-p_t^2),
\end{equation}
\begin{equation}
p_{J/\psi}^*=\frac{1}{2\sqrt{s^*}}\lambda(s^*,m_{J/\psi}^{2},m_{p}^2),
\end{equation}
\begin{equation}
t^+=t(\cos{\theta^*_{J/\psi}}=1)=m_{J/\psi}^2-2E^*_{\gamma}E^*_{J/\psi}+2p^*_{\gamma}p^*_{J/\psi},
\end{equation}
\begin{equation}
t-t^+=2p^*_{\gamma}p^*_{J/\psi}(\cos{\theta^*_{J/\psi}}-1).
\end{equation}
The angle of charmonium production in the ${\gamma}p$ c.m. system, $\theta^*_{J/\psi}$, is expressed
through its production angle, $\theta_{J/\psi}$, in the laboratory system
($\cos{\theta_{J/\psi}}={\bf p}_{\gamma}{\bf p}_{J/\psi}/p_{\gamma}p_{J/\psi}$)
by means of equation [8, 10]:
\begin{equation}
\cos{\theta_{J/\psi}^*}=\frac{p_{\gamma}p_{J/\psi}\cos{\theta_{J/\psi}}+
(E_{\gamma}^*E_{J/\psi}^*-E_{\gamma}E_{J/\psi})}{p_{\gamma}^*p_{J/\psi}^*}.
\end{equation}
The condition of normalization
\begin{equation}
\int\limits_{4\pi}^{}a{\rm e}^{b_{J/\psi}(t-t^+)}d{\bf \Omega}_{J/\psi}^*=1
\end{equation}
gives for the parameter $a$ in Eq. (13) the following expression:
\begin{equation}
a=\frac{p^*_{\gamma}p^*_{J/\psi}b_{J/\psi}}{\pi}\left[1-{\rm e}^{-4p^*_{\gamma}p^*_{J/\psi}b_{J/\psi}}\right]^{-1}.
\end{equation}
Parameter $b_{J/\psi}$ in Eqs. (13), (21), (22)  is an exponential $t$-slope of the differential cross section
of the reaction ${\gamma}p \to {J/\psi}p$ in the near-threshold energy region. It should be pointed out that
the differential cross section of this reaction was also very recently measured in the $J/\psi$-007 experiment [13]
as a function of the photon energy in the range of $9.1~{\rm GeV} \le E_{\gamma} \le 10.6~{\rm GeV}$ with the aim
to explore the impact of the collected data on the determination of the proton's gravitational form factors,
the proton-mass radius, and the contribution of the trace anomaly to the proton mass. Since the $t$-slope
$b_{J/\psi}$ was not determined in [13], we will adopt for it the GlueX result [4], namely:
$b_{J/\psi}$ $\approx$~1.67 GeV$^{-2}$. We will use this value in our calculations.

   Now consider the off-shell total cross section
$\sigma_{{\gamma}p\to {J/\psi}p}(\sqrt{s^*})$ for $J/\psi$ production in process (1).
In line with the above-mentioned, it is the same as the vacuum cross section
$\sigma_{{\gamma}p \to {J/\psi}p}(\sqrt{s})$, in which the vacuum c.m. energy squared s,
defined by the formula (8), is replaced by the in-medium expression (6). For the vacuum total cross section
$\sigma_{{\gamma}p \to {J/\psi}p}(\sqrt{s})$ at near-threshold photon energies
we have used the following parametrization [7] of the available here experimental
information [4] on it from the GlueX experiment, based on the near-threshold predictions
of the two gluon and three gluon exchange model [14]:
\begin{equation}
\sigma_{{\gamma}p \to {J/\psi}p}({\sqrt{s}})= \sigma_{2g}({\sqrt{s}})+
\sigma_{3g}({\sqrt{s}}),
\end{equation}
where 2$g$ and 3$g$ exchanges cross sections $\sigma_{2g}({\sqrt{s}})$ and $\sigma_{3g}({\sqrt{s}})$
are given in Ref. [7] by formulas (7) and (8), respectively.
\begin{figure}[htb]
\begin{center}
\includegraphics[width=16.0cm]{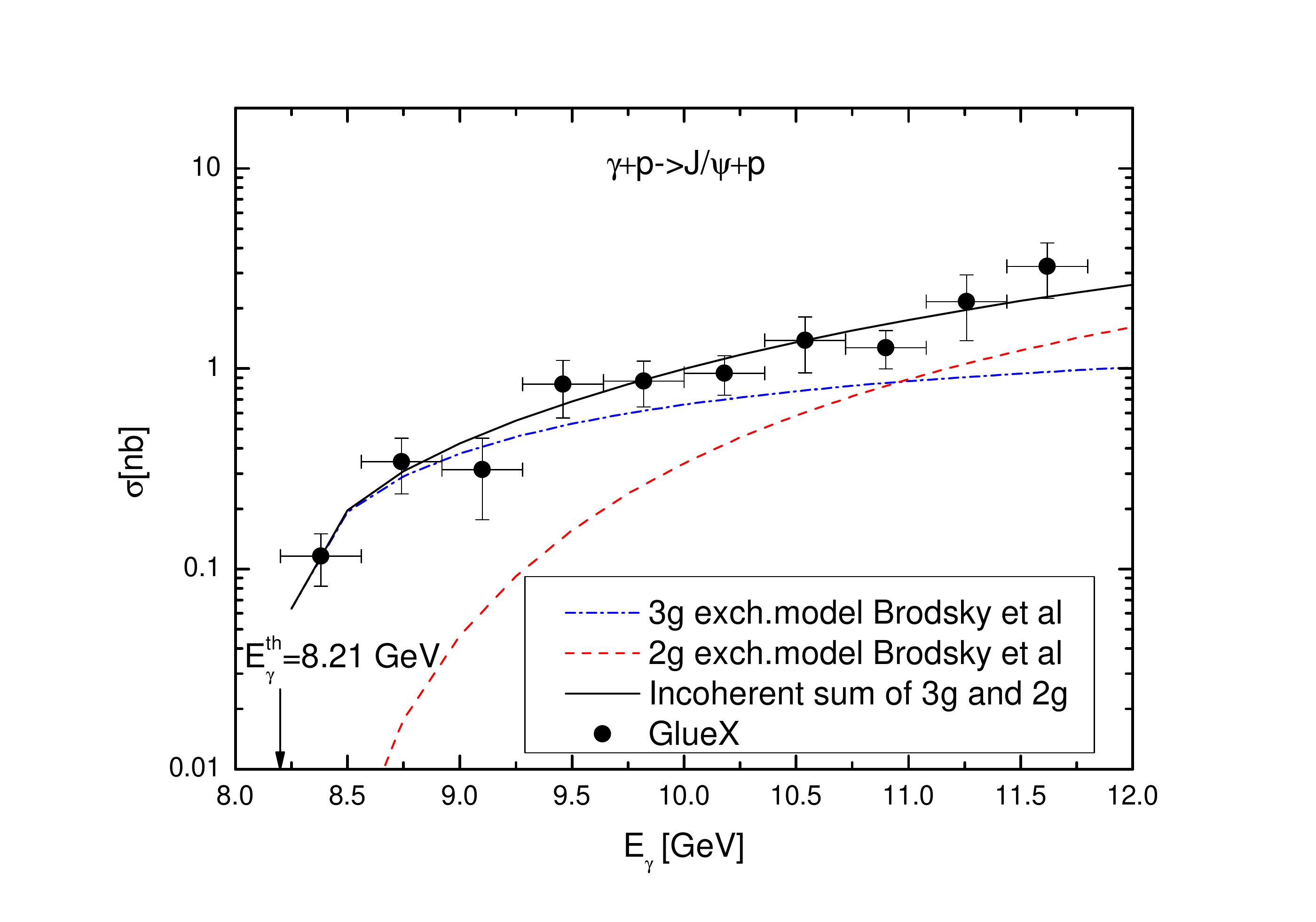}
\vspace*{-2mm} \caption{(Color online) The total cross section for the background reaction
${\gamma}p \to {J/\psi}p$ as a function of the photon energy $E_{\gamma}$.
Dashed and dotted-dashed curves are, respectively, calculations on the basis of the two gluon and
three gluon exchange model [14]. Solid curve is the incoherent sum of the above two calculations.
The GlueX experimental data are from Ref. [4].
The arrow indicates the threshold energy of 8.21 GeV
for direct non-resonant charmonium photoproduction off a free target proton at rest.}
\label{void}
\end{center}
\end{figure}
\begin{figure}[htb]
\begin{center}
\includegraphics[width=16.0cm]{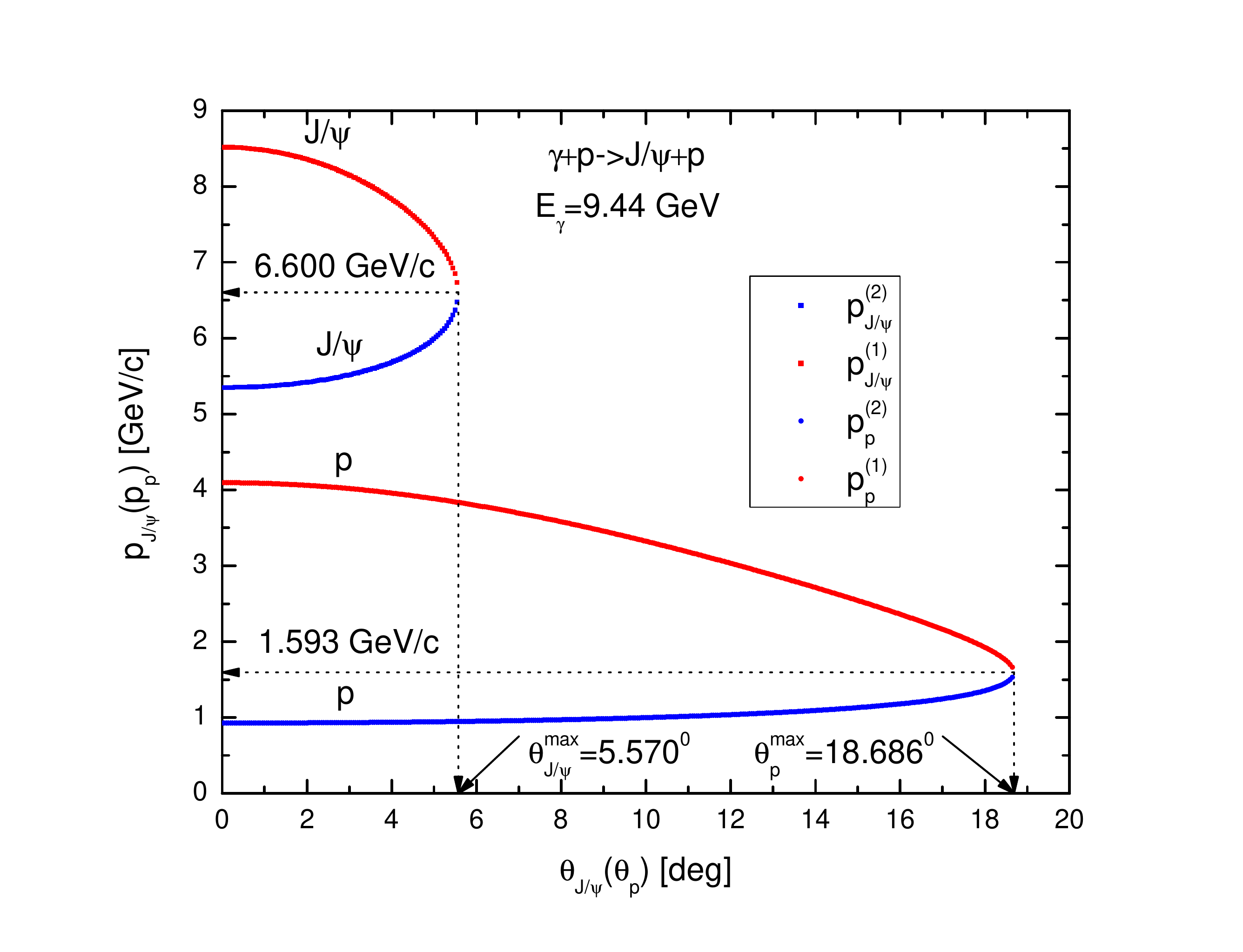}
\vspace*{-2mm} \caption{(Color online) Plot of the allowed final $J/\psi$ meson and proton momenta
in the direct non-resonant ${\gamma}p \to {J/\psi}p$ reaction, occurring
in the laboratory system in the free space at initial photon energy of 9.44 GeV,
as functions of their production angles with respect to the photon beam direction in this system.}
\label{void}
\end{center}
\end{figure}
\begin{figure}[htb]
\begin{center}
\includegraphics[width=16.0cm]{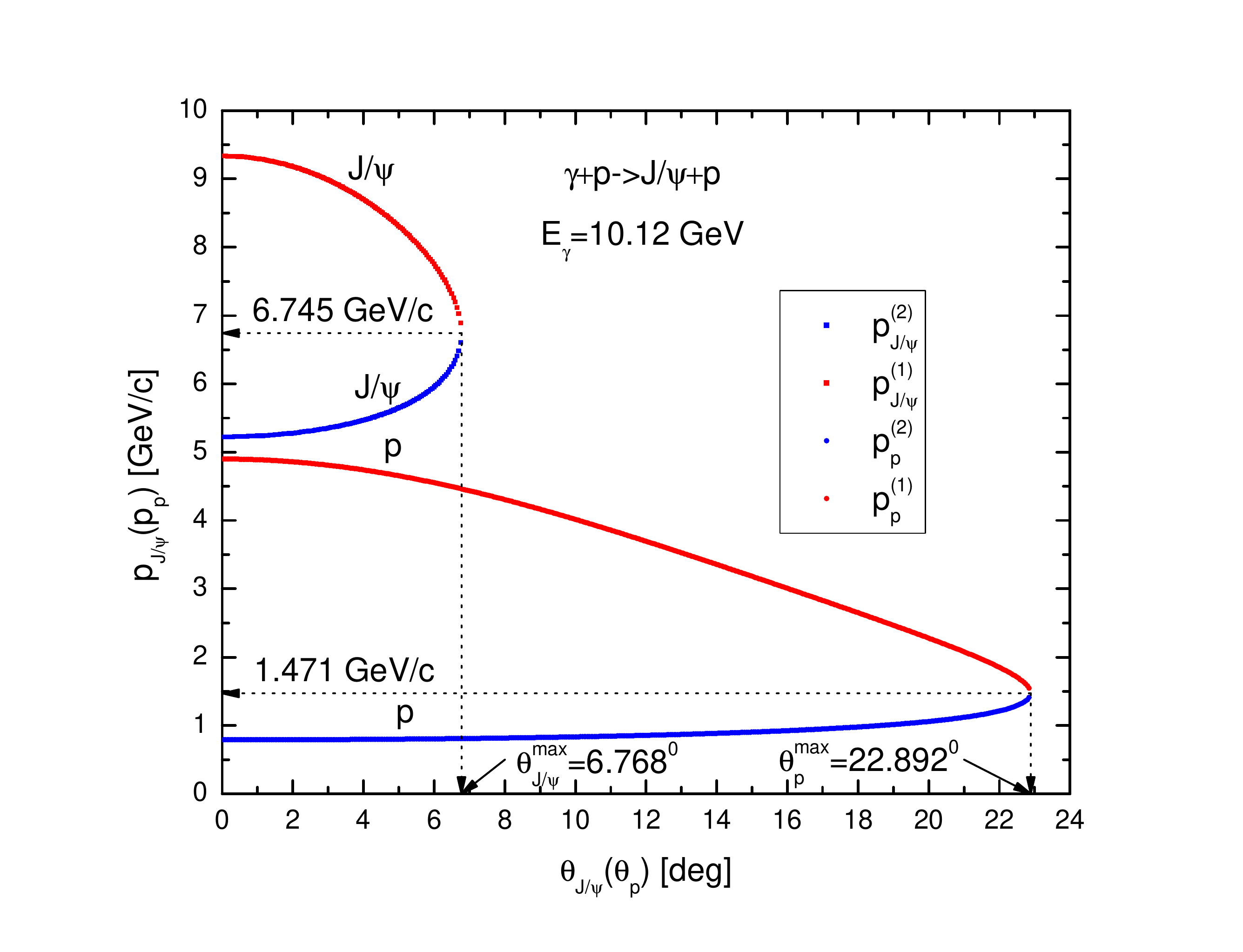}
\vspace*{-2mm} \caption{(Color online) The same as in Fig. 2, but for the initial photon energy of 10.12 GeV.}
\label{void}
\end{center}
\end{figure}
Fig. 1 shows that the GlueX near-threshold data are well fitted by only the combination (23) of the two gluon and
three gluon exchange cross sections and in the resonance incident photon energy range $\sim$ 9.5--10.0 GeV
the main contribution to the elastic $J/\psi$ production comes from the three gluon exchanges.

At the initial photon energies of interest the $J/\psi$ mesons are produced at small
laboratory polar angles (see below). Therefore, we will calculate the $J/\psi$ momentum
distributions from considered target nuclei
for the laboratory solid angle
${\Delta}{\bf \Omega}_{J/\psi}$=$0^{\circ} \le \theta_{J/\psi} \le 20^{\circ}$,
and $0 \le \varphi_{J/\psi} \le 2{\pi}$ (cf. [10]).
Then, integrating the differential cross section (3) over this solid angle,
we can represent the differential cross section for charmonium
production from the direct non-resonant processes (1) and (2) into this solid angle
as follows:
\begin{equation}
\frac{d\sigma_{{\gamma}A\to {J/\psi}X}^{({\rm dir})}
(p_{\gamma},p_{J/\psi})}{dp_{J/\psi}}=
\int\limits_{{\Delta}{\bf \Omega}_{J/\psi}}^{}d{\bf \Omega}_{J/\psi}
\frac{d\sigma_{{\gamma}A\to {J/\psi}X}^{({\rm dir})}
({\bf p}_{\gamma},{\bf p}_{J/\psi})}{d{\bf p}_{J/\psi}}p_{J/\psi}^2
\end{equation}
$$
=2{\pi}I_{V}[A,\sigma_{{J/\psi}N}]
\int\limits_{\cos20^{\circ}}^{1}d\cos{{\theta_{J/\psi}}}
\left<\frac{d\sigma_{{\gamma}p\to {J/\psi}{p}}(p_{\gamma},
p_{J/\psi},\theta_{J/\psi})}{dp_{J/\psi}d{\bf \Omega}_{J/\psi}}\right>_A.
$$

  Before going further, we now consider, adopting the relativistic kinematics,
more simpler case of the free space production of $J/\psi$ mesons and protons
in the process ${\gamma}p \to {J/\psi}p$, proceeding on a free target proton being at rest,
to get an idea about their kinematic characteristics allowed in this process at incident
photon energies considered. The kinematics of two-body reaction with a threshold
(as in our present case) tell us that the laboratory polar $J/\psi$ and final proton
production angles $\theta_{J/\psi}$ and $\theta_{p}$ vary from 0 to a maximal values
$\theta^{\rm max}_{J/\psi}$ and $\theta^{\rm max}_{p}$, correspondingly, i.e.:
\begin{equation}
     0 \le \theta_{J/\psi} \le \theta^{\rm max}_{J/\psi},
\end{equation}
\begin{equation}
     0 \le \theta_{p} \le \theta^{\rm max}_{p};
\end{equation}
where
\begin{equation}
 \theta^{\rm max}_{J/\psi}={\rm arcsin}[(\sqrt{s}p^{*}_{J/\psi})/(m_{J/\psi}p_{\gamma})],
\end{equation}
\begin{equation}
 \theta^{\rm max}_{p}={\rm arcsin}[(\sqrt{s}p^{*}_{p})/(m_{p}p_{\gamma})].
\end{equation}
Here, the $J/\psi$ c.m. momentum $p^*_{J/\psi}$ is determined by Eq. (17),
in which the in-medium c.m. energy squared $s^*$ should be replaced by the
vacuum collision energy squared $s$, defined by the formula (8), and
$p^*_{p}$ is the final proton c.m. momentum. It is equal to the $J/\psi$ c.m. momentum $p^*_{J/\psi}$.
From Eqs. (27), (28) one can get, for example, that
\begin{equation}
\theta^{\rm max}_{J/\psi}=5.570^{\circ},\,\,\,\,\theta^{\rm max}_{p}=18.686^{\circ}
\end{equation}
at initial photon beam energy of $E_{\gamma}=9.44$ GeV and
\begin{equation}
\theta^{\rm max}_{J/\psi}=6.768^{\circ}, \,\,\,\,\theta^{\rm max}_{p}=22.892^{\circ}
\end{equation}
at photon energy of $E_{\gamma}=10.12$ GeV. Energy-momentum conservation in
the reaction (1), taking place in a vacuum, leads to two different solutions for the laboratory
$J/\psi$ meson and final proton momenta $p_{J/\psi}$ and $p_{p}$ at given laboratory polar production angles
$\theta_{J/\psi}$ and $\theta_{p}$, belonging, correspondingly, to the angular intervals (25) and (26):
\begin{equation}
p^{(1,2)}_{J/\psi}(\theta_{J/\psi})=
\frac{p_{\gamma}\sqrt{s}E^{*}_{J/\psi}\cos{\theta_{J/\psi}}\pm
(E_{\gamma}+m_p)\sqrt{s}\sqrt{p^{*2}_{J/\psi}-{\gamma^2_{\rm cm}}{v^2_{\rm cm}}m^2_{J/\psi}\sin^2{\theta_{J/\psi}}}}{(E_{\gamma}+m_p)^2-p^2_{\gamma}\cos^2{\theta_{J/\psi}}},
\end{equation}
\begin{equation}
p^{(1,2)}_{p}(\theta_{p})=
\frac{p_{\gamma}\sqrt{s}E^{*}_{p}\cos{\theta_{p}}\pm
(E_{\gamma}+m_p)\sqrt{s}\sqrt{p^{*2}_{p}-{\gamma^2_{\rm cm}}{v^2_{\rm cm}}m^2_{p}\sin^2{\theta_{p}}}}{(E_{\gamma}+m_p)^2-p^2_{\gamma}\cos^2{\theta_{p}}}.
\end{equation}
Here, ${\gamma_{\rm cm}}=(E_{\gamma}+m_p)/\sqrt{s}$, $v_{\rm cm}=p_{\gamma}/(E_{\gamma}+m_p)$,
the $J/\psi$ total c.m. energy $E^{*}_{J/\psi}$ is defined above by Eq. (15),
$E^{*}_{p}=\sqrt{m^2_{p}+p^{*2}_{p}}$ and
sign "+" in the numerators of Eqs. (31), (32)
corresponds to the first solutions $p^{(1)}_{J/\psi}$, $p^{(1)}_{p}$
and sign "-" - to the second ones $p^{(2)}_{J/\psi}$, $p^{(2)}_{p}$.
Looking at the expressions (31) and (32), one can come to the conclusion that the first solutions $p^{(1)}_{J/\psi}$ and
$p^{(1)}_{p}$ as well as the second ones $p^{(2)}_{J/\psi}$ and $p^{(2)}_{p}$
have different dependencies, respectively, on the production angles $\theta_{J/\psi}$ and $\theta_{p}$
within the angular intervals [0, $\theta_{J/\psi}^{\rm max}]$ and [0, $\theta_{p}^{\rm max}]$.
Namely, the former drop and the latter ones increase as the production angles
$\theta_{J/\psi}$ and $\theta_{p}$ increase in these intervals (cf. Figs. 2 and 3) and
\begin{equation}
p^{(1)}_{J/\psi}(\theta_{J/\psi}^{\rm max})=p^{(2)}_{J/\psi}(\theta_{J/\psi}^{\rm max})=
p_{J/\psi}(\theta_{J/\psi}^{\rm max}),
\end{equation}
\begin{equation}
p^{(1)}_{p}(\theta_{p}^{\rm max})=p^{(2)}_{p}(\theta_{p}^{\rm max})=
p_{p}(\theta_{p}^{\rm max}),
\end{equation}
 where
\begin{equation}
p_{J/\psi}(\theta_{J/\psi}^{\rm max})=(p_{\gamma}m_{J/\psi}^2\cos{\theta_{J/\psi}^{\rm max}})/
(\sqrt{s}E^{*}_{J/\psi}),
\end{equation}
\begin{equation}
p_{p}(\theta_{p}^{\rm max})=(p_{\gamma}m_{p}^2\cos{\theta_{p}^{\rm max}})/
(\sqrt{s}E^{*}_{p}).
\end{equation}
According to Eqs. (35), (36), for $E_{\gamma}=9.44$ GeV we get then that
$p_{J/\psi}(\theta_{J/\psi}^{\rm max})=6.600$ GeV/c and $p_{p}(\theta_{p}^{\rm max})=1.593$ GeV/c.
For $E_{\gamma}=10.12$ GeV we obtain $p_{J/\psi}(\theta_{J/\psi}^{\rm max})=6.745$ GeV/c and
$p_{p}(\theta_{p}^{\rm max})=1.471$ GeV/c (cf. Figs. 2 and 3).
These figures show that the kinematically allowed charmonium laboratory momenta
and total energies in the direct non-resonant ${\gamma}p \to {J/\psi}p$  process, taking place
on the free target proton at rest, at given initial
photon energy vary within the following momentum and energy ranges:
\begin{equation}
p^{(2)}_{J/\psi}(0^{\circ}) \le p_{J/\psi} \le p^{(1)}_{J/\psi}(0^{\circ}),
\end{equation}
\begin{equation}
E^{(2)}_{J/\psi}(0^{\circ}) \le E_{J/\psi} \le E^{(1)}_{J/\psi}(0^{\circ}),
\end{equation}
where the quantities $p^{(1,2)}_{J/\psi}(0^{\circ})$ are defined above
by Eq. (31) and $E^{(1,2)}_{J/\psi}(0^{\circ})=\sqrt{m_{J/\psi}^2+[p^{(1,2)}_{J/\psi}(0^{\circ})]^2}$.

  Finally, we calculate the $J/\psi$ energy distribution
$d\sigma_{{\gamma}p \to {J/\psi}{p}}[\sqrt{s},p_{J/\psi}]/ dE_{J/\psi}$
from the reaction ${\gamma}p \to {J/\psi}p$ within the kinematically allowed interval (38).
Integration of the more general differential cross section (9) over the angle $\theta_{J/\psi}$,
when this angle varies in the allowed angular region (25),
in the limits: ${\bf p}_t \to 0$, $E_t \to m_p$ and $s^* \to s$ yields:
\begin{equation}
\frac{d\sigma_{{\gamma}p \to {J/\psi}{p}}[\sqrt{s},p_{J/\psi}]}{dE_{J/\psi}}=
2\pi
\int\limits_{\cos{\theta_{J/\psi}^{\rm max}}}^{1}d\cos{\theta_{J/\psi}}p_{J/\psi}E_{J/\psi}
\frac{d\sigma_{{\gamma}p\to {J/\psi}p}[\sqrt{s},{\bf p}_{J/\psi}]}{d{\bf p}_{J/\psi}}=
\end{equation}
$$
=
\left(\frac{2{\pi}\sqrt{s}}{p_{\gamma}p^{*}_{J/\psi}}\right)
\frac{d\sigma_{{\gamma}p \to {J/\psi}{p}}[\sqrt{s},\theta_{J/\psi}^*(x_0)]}{d{\bf \Omega}_{J/\psi}^*}~{\rm for}
~E^{(2)}_{J/\psi}(0^{\circ}) \le E_{J/\psi} \le E^{(1)}_{J/\psi}(0^{\circ}),
$$
where
\begin{equation}
x_0=\frac{[p^2_{\gamma}+p^2_{J/\psi}+m^2_{p}-(\omega+m_p)^2]}{2p_{\gamma}p_{J/\psi}},\,\,\,\,\,
p_{J/\psi}=\sqrt{E^2_{J/\psi}-m^2_{J/\psi}}
\end{equation}
and the quantity $\cos{\theta_{J/\psi}^*(x_0)}$ is defined by Eq. (20), in which one has to perform the
replacement: $\cos{\theta_{J/\psi}} \to x_0$, and the photon and $J/\psi$ c.m. momenta $p^*_{\gamma}$
and $p^*_{J/\psi}$ are defined by formulas (16) and (17), correspondingly, in which
one needs also to make the substitutions: $E_t \to m_p$, $p_t \to 0$ and $s^* \to s$.
We will adopt the expression (39) for evaluating the free space $J/\psi$ energy distribution from
the direct process (1), proceeding on a proton at rest, for
incident photon beam resonant energies of 9.44, 9.554, 10.04 and 10.12 GeV (see below).
\begin{figure}[htb]
\begin{center}
\includegraphics[width=16.0cm]{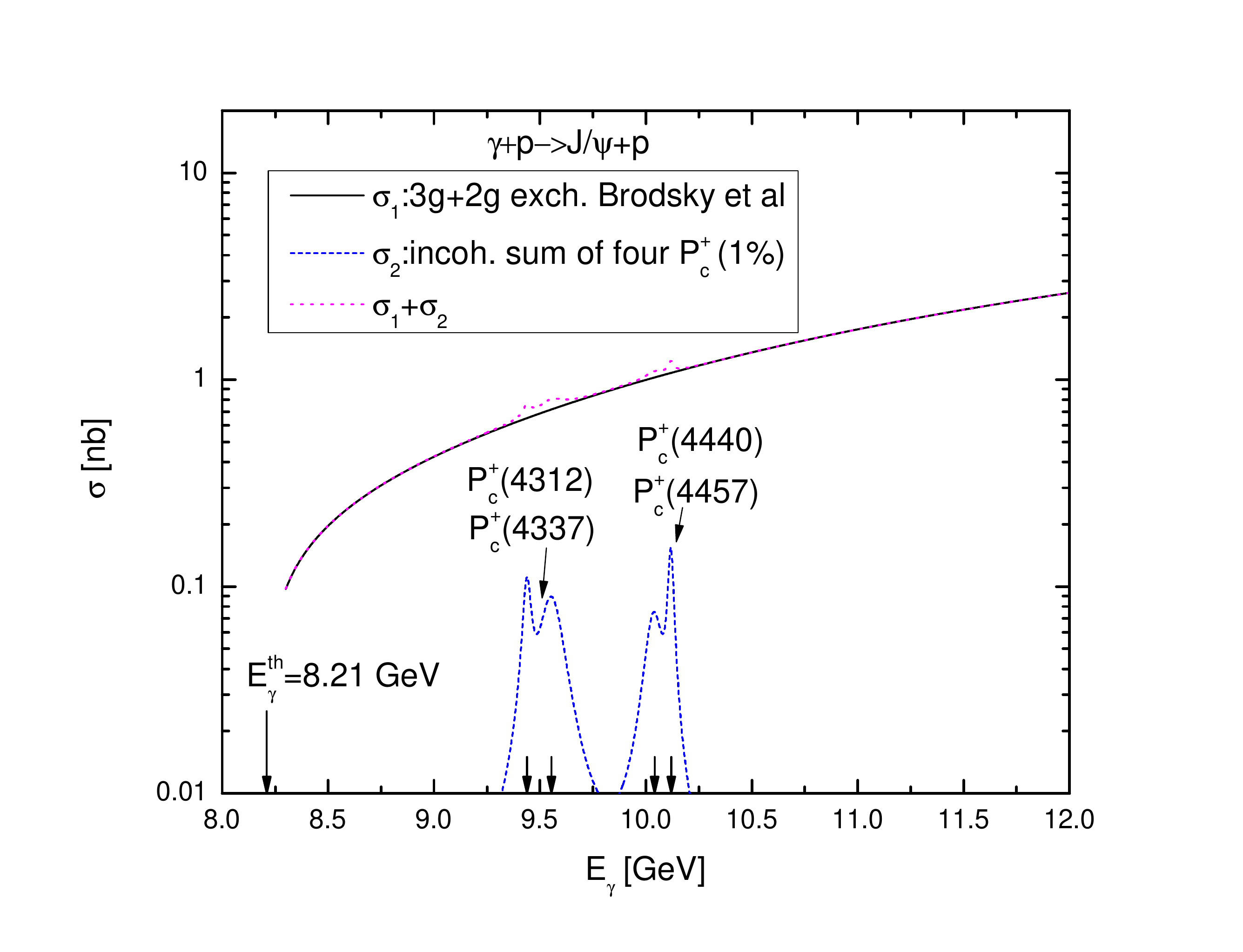}
\vspace*{-2mm} \caption{(Color online) The non-resonant total cross section $\sigma_1$ for the reaction
${\gamma}p \to {J/\psi}p$ (solid curve) and the incoherent sum (dotted curve) of it and the total
cross section $\sigma_2$ (short-dashed curve) for the
resonant $J/\psi$ production in the processes ${\gamma}p \to P^+_c(4312) \to {J/\psi}p$,
${\gamma}p \to P^+_c(4337) \to {J/\psi}p$, ${\gamma}p \to P^+_c(4440) \to {J/\psi}p$
and ${\gamma}p \to P^+_c(4457) \to {J/\psi}p$, calculated assuming that the resonances
$P^+_c(4312)$, $P^+_c(4337)$, $P^+_c(4440)$ and $P^+_c(4457)$ with the spin-parity quantum
numbers $J^P=(1/2)^-$, $J^P=(1/2)^-$, $J^P=(1/2)^-$ and $J^P=(3/2)^-$
decay to ${J/\psi}p$ with the lower allowed relative orbital angular momentum $L=0$
with all four branching fractions $Br[P^+_{ci} \to {J/\psi}p]=$1\%,
as functions of photon energy. The left and four right arrows indicate, correspondingly, the threshold energy
$E^{\rm th}_{\gamma}=8.21$ GeV for the reaction ${\gamma}p \to {J/\psi}p$ proceeding on a free target
proton being at rest and the resonant energies $E^{\rm R1}_{\gamma}=9.44$ GeV, $E^{\rm R2}_{\gamma}=9.554$ GeV,
$E^{\rm R3}_{\gamma}=10.04$ GeV and $E^{\rm R4}_{\gamma}=10.12$ GeV.}
\label{void}
\end{center}
\end{figure}

\subsection*{2.2. Two-step resonant $J/\psi$ production processes}

  At photon energies below 11 GeV, incident photons can produce the observed [1--3] experimentally
non-strange charged $P^+_c(4312)$, $P^+_c(4337)$, $P^+_c(4440)$, $P^+_c(4457)$ pentaquark resonances
with quark structure $|P^+_c>=|uudc{\bar c}>$
and predicted [15], but non-observed yet their neutral isospin partners $P^0_c(4312)$, $P^0_c(4337)$,
$P^0_c(4440)$, $P^0_c(4457)$
\footnote{$^)$The minimal quark content of the $P_c^0$ states is $|P^0_c>=|uddc{\bar c}>$.
Following the observation of the narrow pentaquarks $P^+_c(4312)$, $P^+_c(4440)$ and $P^+_c(4457)$
by the LHCb Collaboration [1, 2], it was proposed to search for the $P_c^0$ states in
${\pi^-}p \to {J/\psi}n$ reaction [16].}$^)$
directly in the first inelastic collisions with intranuclear protons and neutrons
\footnote{$^)$We remind that, for example, the threshold (resonant) energies $E^{\rm R1}_{\gamma}$,
$E^{\rm R2}_{\gamma}$, $E^{\rm R3}_{\gamma}$ and $E^{\rm R4}_{\gamma}$ for the photoproduction
of the $P_c^+$ resonances with pole masses $M_{c1}^+=4311.9$ MeV, $M_{c2}^+=4337.0$ MeV, $M_{c3}^+=4440.3$ MeV
and $M_{c4}^+=4457.3$ MeV [2, 3] on a free target proton being at rest are $E^{\rm R1}_{\gamma}=9.44$ GeV, $E^{\rm R2}_{\gamma}=9.554$ GeV, $E^{\rm R3}_{\gamma}=10.04$ GeV and $E^{\rm R4}_{\gamma}=10.12$ GeV, respectively.}$^)$
:
\begin{eqnarray}
{\gamma}+p \to P^+_c(4312),\nonumber\\
{\gamma}+p \to P^+_c(4337),\nonumber\\
{\gamma}+p \to P^+_c(4440),\nonumber\\
{\gamma}+p \to P^+_c(4457);
\end{eqnarray}
\begin{eqnarray}
{\gamma}+n \to P^0_c(4312),\nonumber\\
{\gamma}+n \to P^0_c(4337),\nonumber\\
{\gamma}+n \to P^0_c(4440),\nonumber\\
{\gamma}+n \to P^0_c(4457).
\end{eqnarray}
Furthermore, the produced pentaquark resonances can decay into the final states ${J/\psi}p$
and ${J/\psi}n$, which will additionally contribute to the $J/\psi$ yield in the ($\gamma$,$J/\psi$) reactions
on protons and nuclei:
\begin{eqnarray}
P^+_c(4312) \to J/\psi+p,\nonumber\\
P^+_c(4337) \to J/\psi+p,\nonumber\\
P^+_c(4440) \to J/\psi+p,\nonumber\\
P^+_c(4457) \to J/\psi+p;
\end{eqnarray}
\begin{eqnarray}
P^0_c(4312) \to J/\psi+n,\nonumber\\
P^0_c(4337) \to J/\psi+n,\nonumber\\
P^0_c(4440) \to J/\psi+n,\nonumber\\
P^0_c(4457) \to J/\psi+n.
\end{eqnarray}
The branching ratios $Br[P^+_{ci} \to {J/\psi}p]$
\footnote{$^)$Here, $i=$1, 2, 3, 4 and $P^{+}_{c1}$, $P^{+}_{c2}$, $P^{+}_{c3}$ and $P^{+}_{c4}$
stand for $P^{+}_c(4312)$, $P^{+}_c(4337)$, $P^{+}_c(4440)$ and $P^{+}_c(4457)$, respectively.
Analogously, $P^{0}_{c1}$, $P^{0}_{c2}$, $P^{0}_{c3}$ and $P^{0}_{c4}$ will denote below
the $P^{0}_c(4312)$, $P^{0}_c(4337)$, $P^{0}_c(4440)$ and $P^{0}_c(4457)$ states.}$^)$
of the decays (43) have not been determined yet.
Model-dependent upper limits on branching fractions $Br[P^+_{c}(4312) \to {J/\psi}p]$,
$Br[P^+_{c}(4440) \to {J/\psi}p]$ and $Br[P^+_{c}(4457) \to {J/\psi}p]$ of several percent
were set by the GlueX Hall-D experiment [4] at JLab, having a moderate statistics (about 470 $J/\psi$ events).
Preliminary results from a factor of 10 more data (about 4000 $J/\psi$ events),
collected in the $J/\psi$--007 Hall-C experiment [17] at JLab as well, focused on the large $t$ region
\footnote{$^)$In which the rather flat resonant production of $J/\psi$ through the $P_c^+$
is expected to be enhanced relative to the suppressed here mostly forward diffractive production.}$^)$
in searching for the LHCb hidden-charm pentaquarks [1, 2], also observe  no signals for them
and will set more stringent upper limits on the above branching fractions and on pentaquark-$J/\psi$
couplings. Based on the branching ratios and
fractions measured by the LHCb and GlueX Collaborations, the authors of Ref. [18] obtain that a lower
limit of $Br[P^+_{c} \to {J/\psi}p]$ is of the order of 0.05\% $\sim$ 0.5\%.
Taking into account these findings, we will adopt in our study for the four branching ratios
$Br[P^+_{ci}\to {J/\psi}p]$ of the decays (43) three following conservative options:
$Br[P^+_{ci} \to {J/\psi}p]=0.25$, 0.5 and 1\% ($i=1,2,3,4$), and in line with Ref. [15], will assume that
$Br[P^0_{ci} \to {J/\psi}n]=Br[P^+_{ci} \to {J/\psi}p]$. This will allow us
to get a better impression of the size of the effect of branching
fractions $Br[P^+_{ci} \to {J/\psi}p]$ and $Br[P^0_{ci} \to {J/\psi}n]$
on the resonant $J/\psi$ yield in ${\gamma}p \to {J/\psi}p$ as well as in
${\gamma}$$^{12}$C $\to {J/\psi}X$ and ${\gamma}$$^{184}$W $\to {J/\psi}X$ reactions.
Moreover, we will also suppose, analogously to [15], for the $P_{ci}^0$ states the same pole masses
$M_{ci}^0$ and total decay width $\Gamma_{ci}^0$ as those $M_{ci}^+$ and $\Gamma_{ci}^+$ for their
hidden-charm charged counterparts $P_{ci}^+$, i.e.: $M_{ci}^0=M_{ci}^+$ and
$\Gamma_{c1}^0=\Gamma_{c1}^+=9.8$ MeV, $\Gamma_{c2}^0=\Gamma_{c2}^+=29.0$ MeV,
$\Gamma_{c3}^0=\Gamma_{c3}^+=20.6$ MeV, $\Gamma_{c4}^0=\Gamma_{c4}^+=6.4$ MeV [2, 3].

In line with Refs. [6, 7, 19], we suppose that the in-medium spectral functions
$S_{ci}^+(\sqrt{s^*},\Gamma_{ci}^+)$ and $S_{ci}^0(\sqrt{s^*},\Gamma_{ci}^0)$
of the intermediate $P_{ci}^{+}$ and $P_{ci}^{0}$ resonances
are described by the non-relativistic Breit-Wigner distributions
\footnote{$^)$We ignore, for reasons of the simplification of calculations, the modification of the
$P_{ci}^{+}$ and $P_{ci}^{0}$ masses and total decay widths in the nuclear matter in our present study.}$^)$
:
\begin{equation}
S_{ci}^+(\sqrt{s^*},\Gamma_{ci}^+)=
\frac{1}{2\pi}\frac{\Gamma_{ci}^+}{(\sqrt{s^*}-M_{ci}^+)^2+({\Gamma}_{ci}^+)^{2}/4}
\end{equation}
and
\begin{equation}
S_{ci}^0(\sqrt{s^*},\Gamma_{ci}^0)
=\frac{1}{2\pi}\frac{\Gamma_{ci}^0}{(\sqrt{s^*}-M_{ci}^0)^2+({\Gamma}_{ci}^0)^{2}/4}.
\end{equation}
The in-medium total cross sections for production of these
resonances with the possible spin-parity quantum numbers $J^P=(1/2)^-$ for $P_{c1}^+$ and $P_{c1}^0$,
$J^P=(1/2)^-$ for $P_{c2}^+$ and $P_{c2}^0$, $J^P=(1/2)^-$ for $P_{c3}^+$ and $P_{c3}^0$, and
$J^P=(3/2)^-$ for $P_{c4}^+$ and $P_{c4}^0$
\footnote{$^)$Which might be assigned to them within
the hadronic molecular scenario for their internal structure (cf. [7, 15, 20--22]).}$^)$
in reactions (41), (42) can be determined, using the spectral functions (45), (46) and known
branching fractions $Br[P_{ci}^+ \to {\gamma}p]$ and $Br[P_{ci}^0 \to {\gamma}n]$ ($i=1$, 2, 3, 4),
as follows [6, 7, 19]:
\begin{equation}
\sigma_{{\gamma}p \to P_{ci}^+}(\sqrt{s^*},\Gamma_{ci}^+)=
f_{ci}\left(\frac{\pi}{p^*_{\gamma}}\right)^2
Br[P_{ci}^+ \to {\gamma}p]S_{ci}^+(\sqrt{s^*},\Gamma_{ci}^+)\Gamma_{ci}^+, \,\,i=1,2,3,4
\end{equation}
and
\begin{equation}
\sigma_{{\gamma}n \to P_{ci}^0}(\sqrt{s^*},\Gamma_{ci}^0)=
f_{ci}\left(\frac{\pi}{p^*_{\gamma}}\right)^2
Br[P_{ci}^0 \to {\gamma}n]S_{ci}^0(\sqrt{s^*},\Gamma_{ci}^0)\Gamma_{ci}^0, \,\,i=1,2,3,4.
\end{equation}
Here, the c.m. 3-momentum in the incoming ${\gamma}N$ channel, $p^*_{\gamma}$,
is defined above by Eq. (16)
\footnote{$^)$For simplicity, we assume that the neutron mass $m_n$ is equal to the proton mass $m_p$.}$^)$
and the ratios of the spin factors $f_{c1}=1$, $f_{c2}=1$,
$f_{c3}=1$, $f_{c4}=2$. Since we are mainly interested in the resonance $P_{c}$ region,
which is not far from the ${J/\psi}N$ production threshold, we suppose
in line with [7, 19, 23] that the hidden-charm pentaquarks $P_{ci}^+$ and
$P_{ci}^0$ decays to ${J/\psi}p$ and ${J/\psi}n$ modes
are dominated by the lowest partial waves with zero relative orbital angular momentum $L$.
In this case, adopting the vector-meson dominance model, one can obtain
that the branching ratios $Br[P_{ci}^0 \to {\gamma}n]$ and $Br[P_{ci}^+ \to {\gamma}p]$
are equal to each other (cf. [24])
\begin{equation}
Br[P_{ci}^0 \to {\gamma}n]=Br[P_{ci}^+ \to {\gamma}p]
\end{equation}
and the latter for $P^+_c(4312)$, $P^+_c(4440)$ and $P^+_c(4457)$
are expressed in the framework of this model via the branching fractions
$Br[P^+_c(4312) \to {J/\psi}p]$, $Br[P^+_c(4440) \to {J/\psi}p]$ and $Br[P^+_c(4457) \to {J/\psi}p]$
by formula (24) from Ref. [7] and within this model we get that
\begin{equation}
Br[P^+_c(4337) \to {\gamma}p]=1.48\cdot10^{-3}Br[P^+_c(4337) \to {J/\psi}p].
\end{equation}
Using Eqs. (47)--(49), we have
\begin{equation}
\sigma_{{\gamma}p \to P_{ci}^+}(\sqrt{s^*},\Gamma_{ci}^+)=\sigma_{{\gamma}n \to P_{ci}^0}(\sqrt{s^*},\Gamma_{ci}^0).
\end{equation}
According to Eq. (47), for example, the free total cross sections
$\sigma_{{\gamma}p \to P_{ci}^+ \to {J/\psi}p}(\sqrt{s},\Gamma_{ci}^+)$
for resonant charmonium production in the two-step processes (41)/(43), taking place on the target proton at rest,
can be represented as follows [6, 7]:
\begin{equation}
\sigma_{{\gamma}p \to P_{ci}^+ \to {J/\psi}p}(\sqrt{s},\Gamma_{ci}^+)=
\sigma_{{\gamma}p \to P_{ci}^+}(\sqrt{s},\Gamma_{ci}^+)\theta[\sqrt{s}-(m_{J/\psi}+m_{p})]
Br[P_{ci}^+ \to {J/\psi}p].
\end{equation}
Here, $\theta(x)$ is the step function and the c.m. 3-momentum in the incoming
${\gamma}p$ channel, $p^*_{\gamma}$, entering into Eq. (47), is determined above by the formula (16),
in which one has to make the replacements $E_t^2-p_t^2 \to m^2_p$ and $s^* \to s$.
In line with Eqs. (47) and (50), we see that these cross sections are proportional to $Br^2[P_{ci}^+ \to {J/\psi}p]$.
This fact enables us to evaluate upper limits on the branching fractions $Br[P_{c}^+(4312) \to {J/\psi}p]$,
$Br[P_{c}^+(4440) \to {J/\psi}p]$ and $Br[P_{c}^+(4457) \to {J/\psi}p]$, which are expected from
preliminary results of the JLab E12-16-007 experiment [17]. According to them, upper limits on the
cross sections (52) for $P_{c}^+(4312)$, $P_{c}^+(4440)$ and $P_{c}^+(4457)$ states almost an order of
magnitude below the respective GlueX limits [4]. With this and within the representation of Eq. (52), we
readily obtain the following relation between upper limits on the above branching fractions, which are
expected from the $J/\psi$-007 experiment, and those already available from the GlueX experiment:
$Br_{J/\psi-007}[P_{ci}^+ \to {J/\psi}p]\approx(1/\sqrt{10})Br_{\rm GlueX}[P_{ci}^+ \to {J/\psi}p]$ ($i=1$, 3, 4).
Model-dependent upper limits on the latter ratios of 4.6\%, 2.3\% and 3.8\% for
$P_{c}^+(4312)$, $P_{c}^+(4440)$ and $P_{c}^+(4457)$, assuming for each $P_{ci}^+$ spin-parity
combination $J^P=(3/2)^-$, were set by the GlueX Collaboration [4]. So that, following the above
relation, we get that $Br_{J/\psi-007}[P_{c}^+(4312) \to {J/\psi}p]\approx$1.46\%,
$Br_{J/\psi-007}[P_{c}^+(4440) \to {J/\psi}p]\approx$0.73\% and
$Br_{J/\psi-007}[P_{c}^+(4457) \to {J/\psi}p]\approx$1.20\%. This means that our choice of 1\%
for upper value of the branching ratios $Br[P_{ci}^+ \to {J/\psi}p]$ for all 4 states is quite
reasonable and justified.

Accounting for the fact that the most of the narrow $P_{ci}^+$ amd $P_{ci}^0$ resonances ($i=1$, 2, 3, 4),
having vacuum total decay widths in their rest frames of
9.8, 29.0, 20.6 and 6.4 MeV [2, 3], respectively, decay to ${J/\psi}p$ and ${J/\psi}n$
outside of the considered target nuclei [6] as well as
the results presented both in Refs. [6, 7, 10] and above by
Eqs. (3), (4), (51), (52), we can obtain the following expression for the $J/\psi$
inclusive differential cross section arising from the production and
decay of intermediate resonances $P_{ci}^+$ and $P_{ci}^0$ in ${\gamma}A$ reactions:
\begin{equation}
\frac{d\sigma_{{\gamma}A \to {J/\psi}X}^{({\rm sec})}({\bf p}_{\gamma},{\bf p}_{J/\psi})}{d{\bf p}_{J/\psi}}=
\frac{d\sigma_{{\gamma}A \to P_{ci}^+ \to {J/\psi}p}^{({\rm sec})}({\bf p}_{\gamma},{\bf p}_{J/\psi})}{d{\bf p}_{J/\psi}}
+
\frac{d\sigma_{{\gamma}A \to P_{ci}^0 \to {J/\psi}n}^{({\rm sec})}({\bf p}_{\gamma},{\bf p}_{J/\psi})}{d{\bf p}_{J/\psi}}
=
\end{equation}
$$
=
I_{V}[A,\sigma^{\rm in}_{{P_{c}}N}]
\left<\frac{d\sigma_{{\gamma}p \to P_{ci}^+ \to {J/\psi}p}({\bf p}_{\gamma},{\bf p}_{J/\psi})}
{d{\bf p}_{J/\psi}}\right>_A, i=1, 2, 3, 4,
$$
where
\begin{equation}
\left<\frac{d\sigma_{{\gamma}p \to P_{ci}^+ \to {J/\psi}p}({\bf p}_{\gamma},{\bf p}_{J/\psi})}
{d{\bf p}_{J/\psi}}\right>_A=
\int\int P_A({\bf p}_t,E)d{\bf p}_tdE
\left[\frac{d\sigma_{{\gamma}p \to P_{ci}^+ \to {J/\psi}p}(\sqrt{s^*},{\bf p}_{J/\psi})}
{d{\bf p}_{J/\psi}}\right],
\end{equation}
and
\begin{equation}
\frac{d\sigma_{{\gamma}p \to P_{ci}^+ \to {J/\psi}p}(\sqrt{s^*},{\bf p}_{J/\psi})}
{d{\bf p}_{J/\psi}}=\sigma_{{\gamma}p \to P_{ci}^+}(\sqrt{s^*},\Gamma_{ci}^+)
\theta[\sqrt{s^*}-(m_{J/\psi}+m_{p})]\times
\end{equation}
$$
\times
\frac{1}{\Gamma_{ci}^+(\sqrt{s^*},{\bf p}_{\gamma})}\int d{\bf p}_{p}
\frac{d\Gamma_{P_{ci}^+ \to {J/\psi}p}(\sqrt{s^*},{\bf p}_{J/\psi},{\bf p}_{p})}
{d{\bf p}_{J/\psi}d{\bf p}_{p}},
$$
\begin{equation}
\frac{d\Gamma_{P_{ci}^+ \to {J/\psi}p}(\sqrt{s^*},{\bf p}_{J/\psi},{\bf p}_{p})}
{d{\bf p}_{J/\psi}d{\bf p}_{p}}=\frac{1}{2E_{ci}^+}\frac{1}{2J+1}|M_{P_{ci}^+ \to {J/\psi}p}|^2
(2\pi)^4\delta(E_{ci}^+-E_{J/\psi}-E_{p})\times
\end{equation}
$$
\times
\delta({\bf p}_{ci}^+-{\bf p}_{J/\psi}-{\bf p}_{p})\frac{1}{(2\pi)^3{2E_{J/\psi}}}
\frac{1}{(2\pi)^3{2E_{p}}},
$$
\begin{equation}
\Gamma_{ci}^+(\sqrt{s^*},{\bf p}_{\gamma})=\Gamma_{ci}^+/\gamma_{ci}^+,
\end{equation}
\begin{equation}
E_{ci}^+=E_{\gamma}+E_t,\,\,\,\,\,{\bf p}_{ci}^+={\bf p}_{\gamma}+
{\bf p}_{t},\,\,\,\,\,\gamma_{ci}^+=E_{ci}^+/\sqrt{s^*}.
\end{equation}
Here, $E_{p}$ is the final proton total energy ($E_{p}=\sqrt{m^2_{p}+{\bf p}^2_{p}}$)
and $|M_{P_{ci}^+ \to {J/\psi}p}|^2$ is summarized over spin states of initial and final particles
matrix element squared describing the decays (43) for given $i$.
The quantity $I_{V}[A,\sigma^{\rm in}_{P_{c}N}]$ in Eq. (53) is defined above by Eq. (4), in which one
needs to make the substitution $\sigma \to \sigma^{\rm in}_{P_{c}N}$. Here the quantity
$\sigma^{\rm in}_{P_{c}N}$ denotes the inelastic total cross sections of the free $P_{c}N$ interaction.
Our estimates [6]
\footnote{$^)$These estimates also show that we can neglect quasielastic $P_{ci}^+{N}$ and
$P_{ci}^0{N}$ rescatterings in their way out of the target nucleus.}$^)$,
based on the ${J/\psi}p$ molecular scenario for the $P_{c}^+$ pentaquarks,
show that this quantity can be evaluated as $\sigma^{\rm in}_{P_{c}N} \approx 33.5$ mb. We will use
this value throughout our calculations.
In view of the aforesaid, the hidden-charm pentaquarks $P_{ci}^+$
(and $P_{ci}^0$) decays to ${J/\psi}p$ (and ${J/\psi}n$)
are dominated by the lowest partial $s$-waves with zero relative orbital angular momentum.
This implies that the matrix elements squared
$|M_{P_{ci}^+ \to {J/\psi}p}|^2$ (and $|M_{P_{ci}^0 \to {J/\psi}n}|^2$)
lead to an isotropic angular distributions of the
$P_{ci}^+ \to {J/\psi}p$ (and $P_{ci}^0 \to {J/\psi}n$)
decays for the considered spin-parity assignments of
the $P_{ci}^+$ (and $P_{ci}^0$) states. With this, we can readily
obtain the following relation between $|M_{P_{ci}^+ \to {J/\psi}p}|^2$ and the partial
width $\Gamma_{P_{ci}^+ \to {J/\psi}p}$ of the $P_{ci}^+ \to {J/\psi}p$ decay (cf. [10]):
\begin{equation}
\frac{1}{2J+1}\frac{|M_{P_{ci}^+ \to {J/\psi}p}|^2}{(2\pi)^2}=
\frac{2s^*}{\pi{p^*_{J/\psi}}}\Gamma_{P_{ci}^+ \to {J/\psi}p}.
\end{equation}
With it, we find for the expression (55) a more simpler form (cf. Eq. (9)):
\begin{equation}
\frac{d\sigma_{{\gamma}p \to P_{ci}^+ \to {J/\psi}p}(\sqrt{s^*},{\bf p}_{J/\psi})}
{d{\bf p}_{J/\psi}}=\sigma_{{\gamma}p \to P_{ci}^+}(\sqrt{s^*},\Gamma_{ci}^+)
\theta[\sqrt{s^*}-(m_{J/\psi}+m_{p})]\times
\end{equation}
$$
\times
\frac{1}{I_2(s^*,m_{J/\psi},m_{p})}Br[P_{ci}^+ \to {J/\psi}p]
\frac{1}{4E_{J/\psi}}\frac{1}{(\omega+E_t)}
\delta\left[\omega+E_t-\sqrt{m_{p}^2+({\bf Q}+{\bf p}_t)^2}\right],
$$
where the quantities $\omega$ and ${\bf Q}$ are defined above by Eq. (12).
We will employ this expression in our calculations of the $J/\psi$ momentum distribution from the processes
(41)--(44) in ${\gamma}A$ reactions. Integrating the differential cross section (53)
over the angular range of ${\Delta}{\bf \Omega}_{J/\psi}$=$0^{\circ} \le \theta_{J/\psi} \le 20^{\circ}$,
$0 \le \varphi_{J/\psi} \le 2{\pi}$ of our interest,
we represent this distribution for given $i$ in this angular range in the following form (cf. Eq. (24)):
\begin{equation}
\frac{d\sigma_{{\gamma}A\to {J/\psi}X}^{({\rm sec})}
(p_{\gamma},p_{J/\psi})}{dp_{J/\psi}}=
\int\limits_{{\Delta}{\bf \Omega}_{J/\psi}}^{}d{\bf \Omega}_{J/\psi}
\frac{d\sigma_{{\gamma}A\to {J/\psi}X}^{({\rm sec})}
({\bf p}_{\gamma},{\bf p}_{J/\psi})}{d{\bf p}_{J/\psi}}p_{J/\psi}^2
\end{equation}
$$
=2{\pi}I_{V}[A,\sigma^{\rm in}_{{P_{c}}N}]
\int\limits_{\cos20^{\circ}}^{1}d\cos{{\theta_{J/\psi}}}
\left<\frac{d\sigma_{{\gamma}p \to P_{ci}^+ \to {J/\psi}{p}}(p_{\gamma},
p_{J/\psi},\theta_{J/\psi})}{dp_{J/\psi}d{\bf \Omega}_{J/\psi}}\right>_A,\,\,\,i=1, 2, 3, 4.
$$

Before going to the next step, we calculate the free space resonant $J/\psi$ energy distribution
$d\sigma_{{\gamma}p \to P_{ci}^+ \to {J/\psi}{p}}[\sqrt{s},p_{J/\psi}]/ dE_{J/\psi}$
from the two-step processes (41)/(43), proceeding on the free target proton at rest,
in addition to that from the
background ${\gamma}p \to {J/\psi}p$ reaction (cf. Eq. (39)).
The energy-momentum conservation in these precesses
leads to the conclusion that the kinematical characteristics
of $J/\psi$ mesons produced in them and in this reaction are the same at given incident photon energy.
The full on-shell differential cross section
$d\sigma_{{\gamma}p \to P_{ci}^+ \to {J/\psi}{p}}[\sqrt{s},{\bf p}_{J/\psi}]/ d{\bf p}_{J/\psi}$
can be obtained from more general one (60) in the limits: ${\bf p}_t \to 0$,
$E_t \to m_p$ and $s^* \to s$. Its integration over the laboratory polar angle $\theta_{J/\psi}$,
when this angle belongs to the allowed angular interval (25), gives:
\begin{equation}
\frac{d\sigma_{{\gamma}p \to P_{ci}^+ \to {J/\psi}{p}}[\sqrt{s},p_{J/\psi}]}{dE_{J/\psi}}=
2\pi
\int\limits_{\cos{\theta_{J/\psi}^{\rm max}}}^{1}d\cos{\theta_{J/\psi}}p_{J/\psi}E_{J/\psi}
\frac{d\sigma_{{\gamma}p \to P_{ci}^+ \to {J/\psi}p}[\sqrt{s},{\bf p}_{J/\psi}]}{d{\bf p}_{J/\psi}}=
\end{equation}
$$
=
\sigma_{{\gamma}p \to P_{ci}^+}(\sqrt{s},\Gamma_{ci}^+)\theta[\sqrt{s}-(m_{J/\psi}+m_{p})]\times
$$
$$
\times
\left(\frac{\sqrt{s}}{2p_{\gamma}p^{*}_{J/\psi}}\right)
Br[P_{ci}^+ \to {J/\psi}p]~{\rm for}
~E^{(2)}_{J/\psi}(0^{\circ}) \le E_{J/\psi} \le E^{(1)}_{J/\psi}(0^{\circ}).
$$
Eq. (62) shows that the free space $J/\psi$ energy distribution, which arises from the
production/decay chains (41)/(43), exhibits a completely flat behavior within the
allowed energy range (38).
\begin{figure}[!h]
\begin{center}
\includegraphics[width=16.0cm]{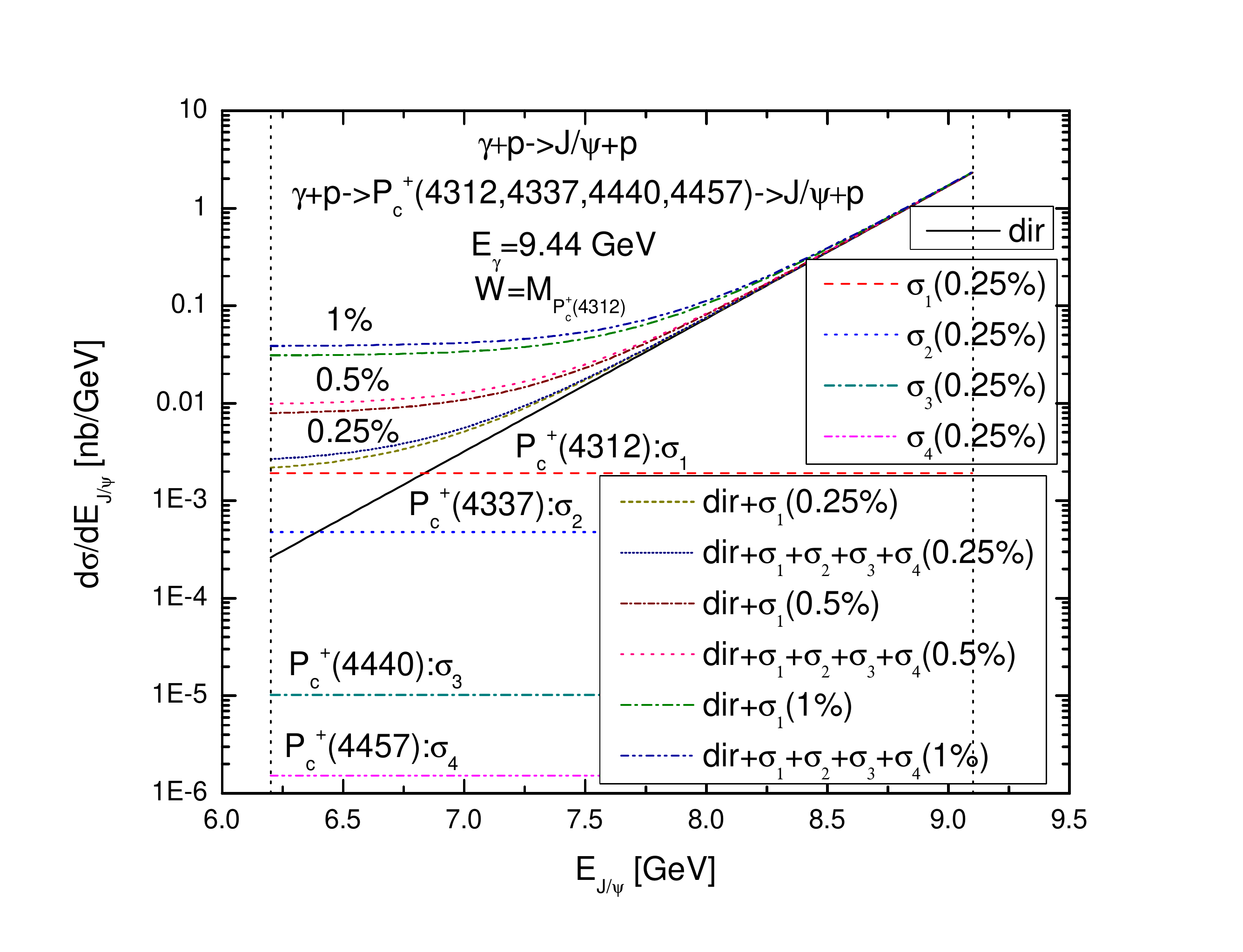}
\vspace*{-2mm} \caption{(Color online) The direct non-resonant $J/\psi$ energy distribution in the free space
elementary process ${\gamma}p \to {J/\psi}p$,
calculated in line with Eq. (39) at initial photon resonant energy of 9.44 GeV
in the laboratory system (solid curve). The resonant $J/\psi$ energy distributions in the two-step processes
${\gamma}p \to P_{c}^+(4312) \to {J/\psi}p$, ${\gamma}p \to P_{c}^+(4337) \to {J/\psi}p$,
${\gamma}p \to P_{c}^+(4440) \to {J/\psi}p$ and ${\gamma}p \to P_{c}^+(4457) \to {J/\psi}p$,
calculated in line with Eq. (62) at the same incident photon energy of 9.44 GeV
assuming that the resonances $P_{c}^+(4312)$, $P_{c}^+(4337)$, $P_{c}^+(4440)$ and $P_{c}^+(4457)$
with the spin-parity assignments $J^P=(1/2)^-$, $J^P=(1/2)^-$, $J^P=(1/2)^-$ and $J^P=(3/2)^-$, correspondingly,
all decay to the ${J/\psi}p$
with branching fractions 0.25\% (respectively, red dashed, blue dotted, dark cyan dashed-doted
and magenta dashed-dotted-dotted curves).
Incoherent sum of the direct non-resonant $J/\psi$ energy distribution and resonant ones, calculated supposing that the resonances $P_{c}^+(4312)$ and $P_{c}^+(4312)$, $P_{c}^+(4337)$, $P_{c}^+(4440)$, $P_{c}^+(4457)$
with the same spin-parity combinations all decay to the ${J/\psi}p$ with branching
fractions 0.25, 0.5 and 1\% (respectively, dark yellow short-dashed, wine short-dashed-dotted, olive dashed-dotted and navy short-dotted, pink dotted, royal dashed-dotted-dotted curves), all as functions of the total $J/\psi$ energy $E_{J/\psi}$ in the laboratory system.
The vertical dotted lines indicate the range of $J/\psi$ allowed energies in this system
for the considered direct non-resonant and resonant $J/\psi$ production off a free target proton at rest at
given initial photon resonant energy of 9.44 GeV.}
\label{void}
\end{center}
\end{figure}
\begin{figure}[!h]
\begin{center}
\includegraphics[width=16.0cm]{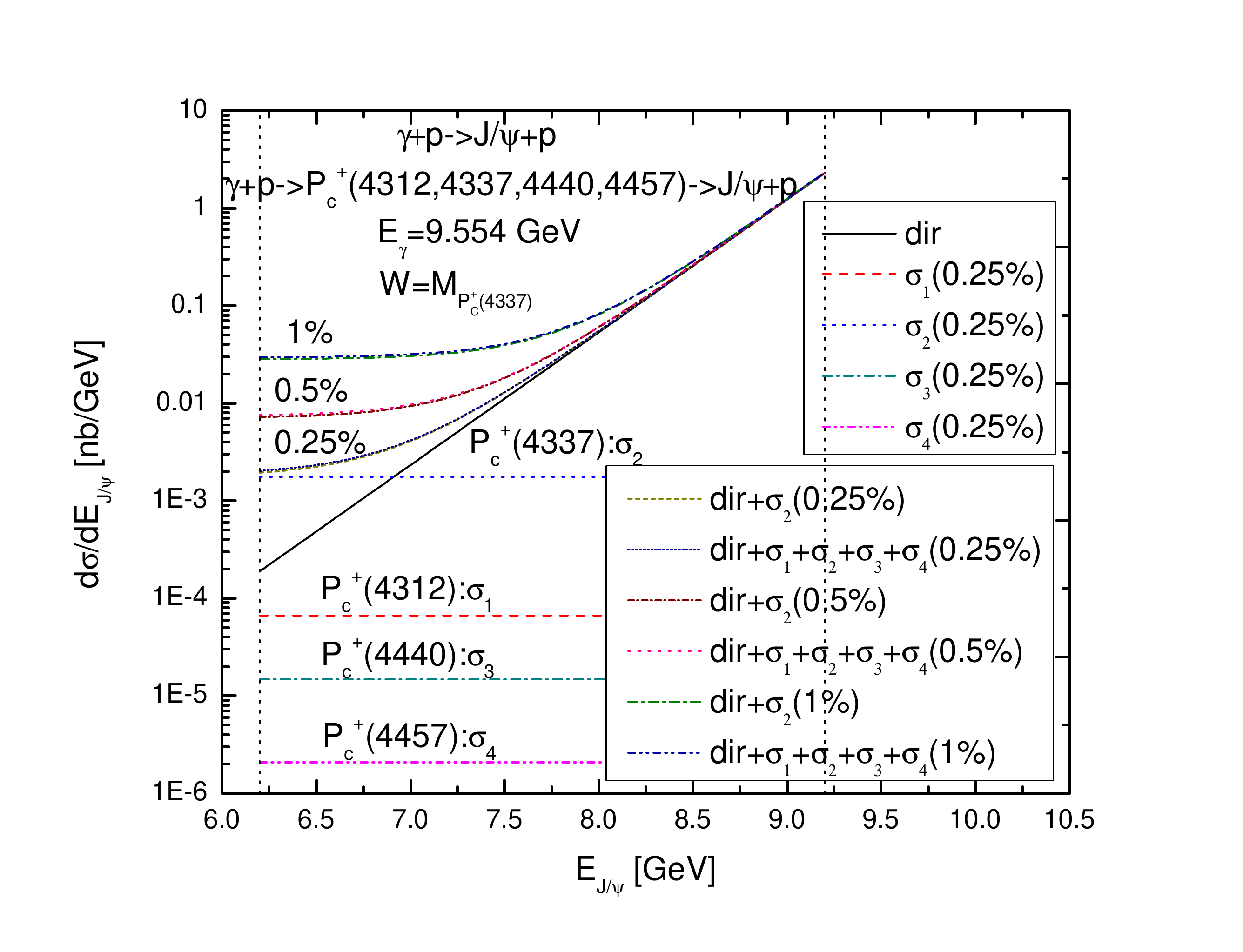}
\vspace*{-2mm} \caption{(Color online) The direct non-resonant $J/\psi$ energy distribution in the free space
elementary process ${\gamma}p \to {J/\psi}p$,
calculated in line with Eq. (39) at initial photon resonant energy of 9.554 GeV
in the laboratory system (solid curve). The resonant $J/\psi$ energy distributions in the two-step processes
${\gamma}p \to P_{c}^+(4312) \to {J/\psi}p$, ${\gamma}p \to P_{c}^+(4337) \to {J/\psi}p$,
${\gamma}p \to P_{c}^+(4440) \to {J/\psi}p$ and ${\gamma}p \to P_{c}^+(4457) \to {J/\psi}p$,
calculated in line with Eq. (62) at the same incident photon energy of 9.554 GeV
assuming that the resonances $P_{c}^+(4312)$, $P_{c}^+(4337)$, $P_{c}^+(4440)$ and $P_{c}^+(4457)$
with the spin-parity assignments $J^P=(1/2)^-$, $J^P=(1/2)^-$, $J^P=(1/2)^-$ and $J^P=(3/2)^-$, correspondingly,
all decay to the ${J/\psi}p$
with branching fractions 0.25\% (respectively, red dashed, blue dotted, dark cyan dashed-doted
and magenta dashed-dotted-dotted curves).
Incoherent sum of the direct non-resonant $J/\psi$ energy distribution and resonant ones, calculated supposing that the resonances $P_{c}^+(4337)$ and $P_{c}^+(4312)$, $P_{c}^+(4337)$, $P_{c}^+(4440)$, $P_{c}^+(4457)$
with the same spin-parity combinations
all decay to the ${J/\psi}p$ with branching
fractions 0.25, 0.5 and 1\% (respectively, dark yellow short-dashed, wine short-dashed-dotted, olive dashed-dotted and navy short-dotted, pink dotted, royal dashed-dotted-dotted curves), all as functions of the total $J/\psi$ energy $E_{J/\psi}$ in the laboratory system.
The vertical dotted lines indicate the range of $J/\psi$ allowed energies in this system
for the considered direct non-resonant and resonant $J/\psi$ production off a free target proton at rest at
given initial photon resonant energy of 9.554 GeV.}
\label{void}
\end{center}
\end{figure}
\begin{figure}[!h]
\begin{center}
\includegraphics[width=16.0cm]{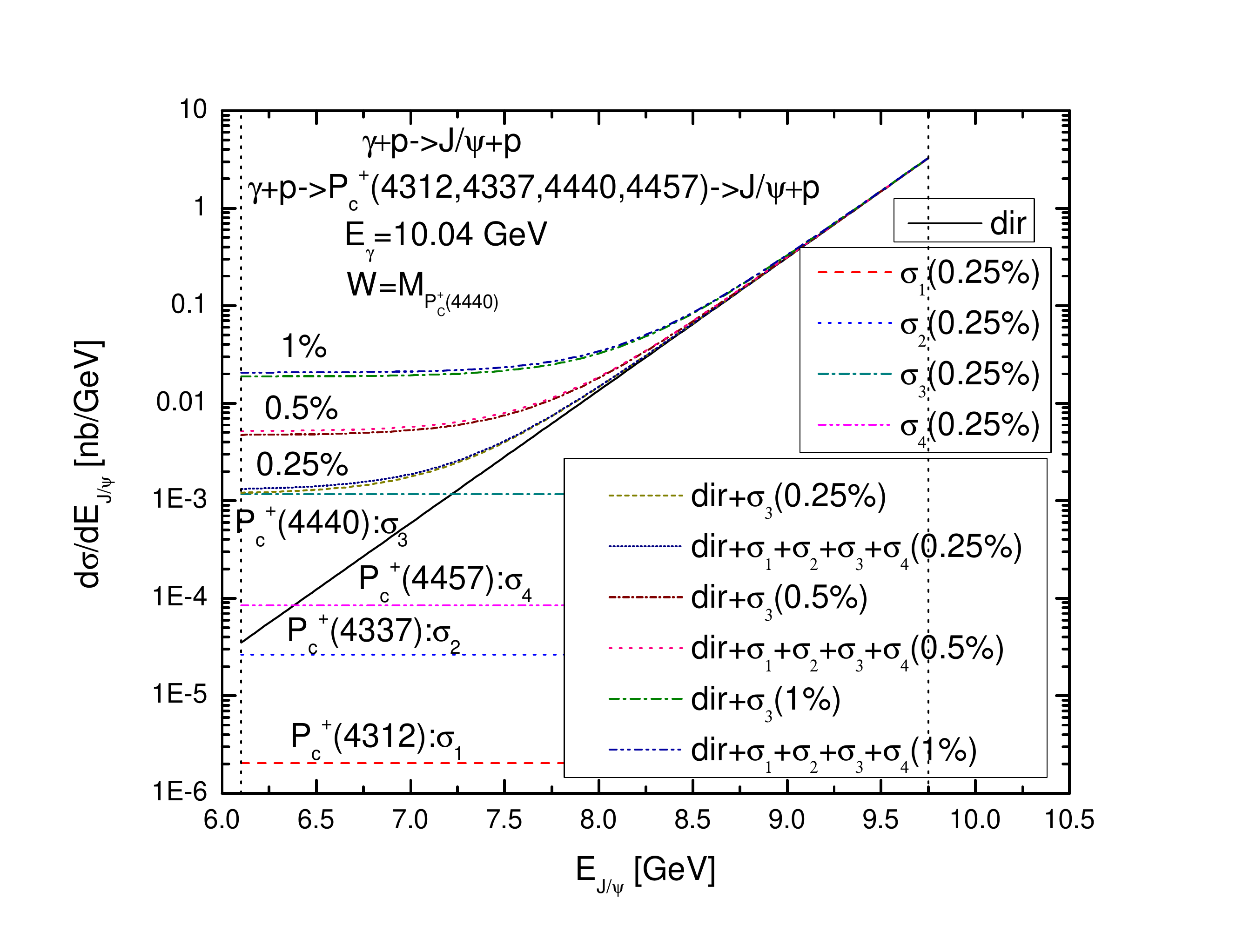}
\vspace*{-2mm} \caption{(Color online) The direct non-resonant $J/\psi$ energy distribution in the free space
elementary process ${\gamma}p \to {J/\psi}p$,
calculated in line with Eq. (39) at initial photon resonant energy of 10.04 GeV
in the laboratory system (solid curve). The resonant $J/\psi$ energy distributions in the two-step processes
${\gamma}p \to P_{c}^+(4312) \to {J/\psi}p$, ${\gamma}p \to P_{c}^+(4337) \to {J/\psi}p$,
${\gamma}p \to P_{c}^+(4440) \to {J/\psi}p$ and ${\gamma}p \to P_{c}^+(4457) \to {J/\psi}p$,
calculated in line with Eq. (62) at the same incident photon energy of 10.04 GeV
assuming that the resonances $P_{c}^+(4312)$, $P_{c}^+(4337)$, $P_{c}^+(4440)$ and $P_{c}^+(4457)$
with the spin-parity assignments $J^P=(1/2)^-$, $J^P=(1/2)^-$, $J^P=(1/2)^-$ and $J^P=(3/2)^-$, correspondingly,
all decay to the ${J/\psi}p$
with branching fractions 0.25\% (respectively, red dashed, blue dotted, dark cyan dashed-doted
and magenta dashed-dotted-dotted curves).
Incoherent sum of the direct non-resonant $J/\psi$ energy distribution and resonant ones, calculated supposing that the resonances $P_{c}^+(4440)$ and $P_{c}^+(4312)$, $P_{c}^+(4337)$, $P_{c}^+(4440)$, $P_{c}^+(4457)$
with the same spin-parity combinations all decay to the ${J/\psi}p$ with branching
fractions 0.25, 0.5 and 1\% (respectively, dark yellow short-dashed, wine short-dashed-dotted, olive dashed-dotted and navy short-dotted, pink dotted, royal dashed-dotted-dotted curves), all as functions of the total $J/\psi$ energy $E_{J/\psi}$ in the laboratory system.
The vertical dotted lines indicate the range of $J/\psi$ allowed energies
in this system for the considered direct non-resonant and resonant $J/\psi$ production
off a free target proton at rest at given initial photon resonant energy of 10.04 GeV.}
\label{void}
\end{center}
\end{figure}
\begin{figure}[!h]
\begin{center}
\includegraphics[width=16.0cm]{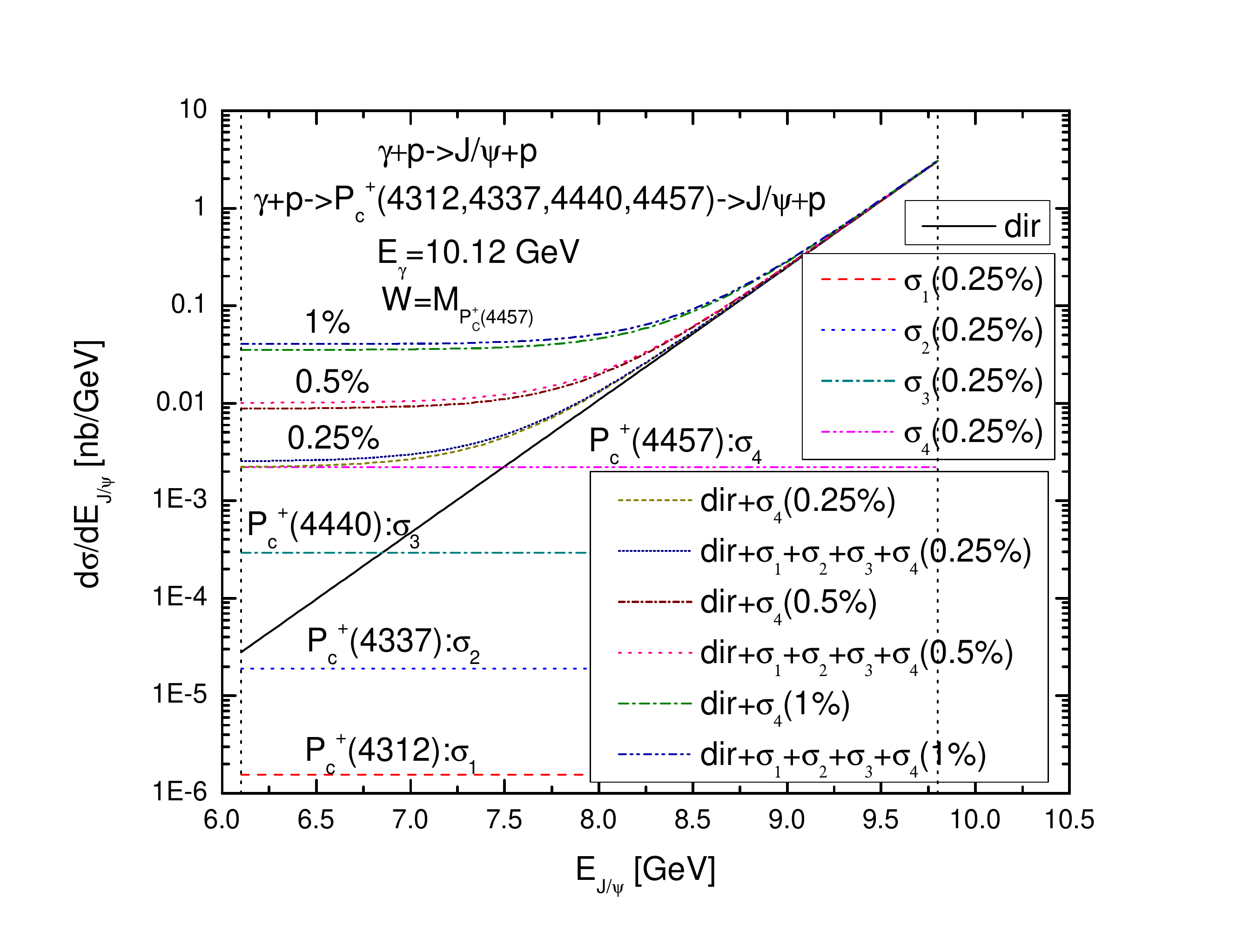}
\vspace*{-2mm} \caption{(Color online) The direct non-resonant $J/\psi$ energy distribution in the free space
elementary process ${\gamma}p \to {J/\psi}p$,
calculated in line with Eq. (39) at initial photon resonant energy of 10.12 GeV
in the laboratory system (solid curve). The resonant $J/\psi$ energy distributions in the two-step processes
${\gamma}p \to P_{c}^+(4312) \to {J/\psi}p$, ${\gamma}p \to P_{c}^+(4337) \to {J/\psi}p$,
${\gamma}p \to P_{c}^+(4440) \to {J/\psi}p$ and ${\gamma}p \to P_{c}^+(4457) \to {J/\psi}p$,
calculated in line with Eq. (62) at the same incident photon energy of 10.12 GeV
assuming that the resonances $P_{c}^+(4312)$, $P_{c}^+(4337)$, $P_{c}^+(4440)$ and $P_{c}^+(4457)$
with the spin-parity assignments $J^P=(1/2)^-$, $J^P=(1/2)^-$, $J^P=(1/2)^-$ and $J^P=(3/2)^-$, correspondingly,
all decay to the ${J/\psi}p$
with branching fractions 0.25\% (respectively, red dashed, blue dotted, dark cyan dashed-doted
and magenta dashed-dotted-dotted curves).
Incoherent sum of the direct non-resonant $J/\psi$ energy distribution and resonant ones, calculated supposing that the resonances $P_{c}^+(4457)$ and $P_{c}^+(4312)$, $P_{c}^+(4337)$, $P_{c}^+(4440)$, $P_{c}^+(4457)$
with the same spin-parity combinations all decay to the ${J/\psi}p$ with branching
fractions 0.25, 0.5 and 1\% (respectively, dark yellow short-dashed, wine short-dashed-dotted, olive dashed-dotted and navy short-dotted, pink dotted, royal dashed-dotted-dotted curves), all as functions of the total $J/\psi$ energy $E_{J/\psi}$ in the laboratory system.
The vertical dotted lines indicate the range of $J/\psi$ allowed energies in this system for the considered direct non-resonant and resonant $J/\psi$ production off a free target proton at rest at
given initial photon resonant energy of 10.12 GeV.}
\label{void}
\end{center}
\end{figure}
\begin{figure}[!h]
\begin{center}
\includegraphics[width=16.0cm]{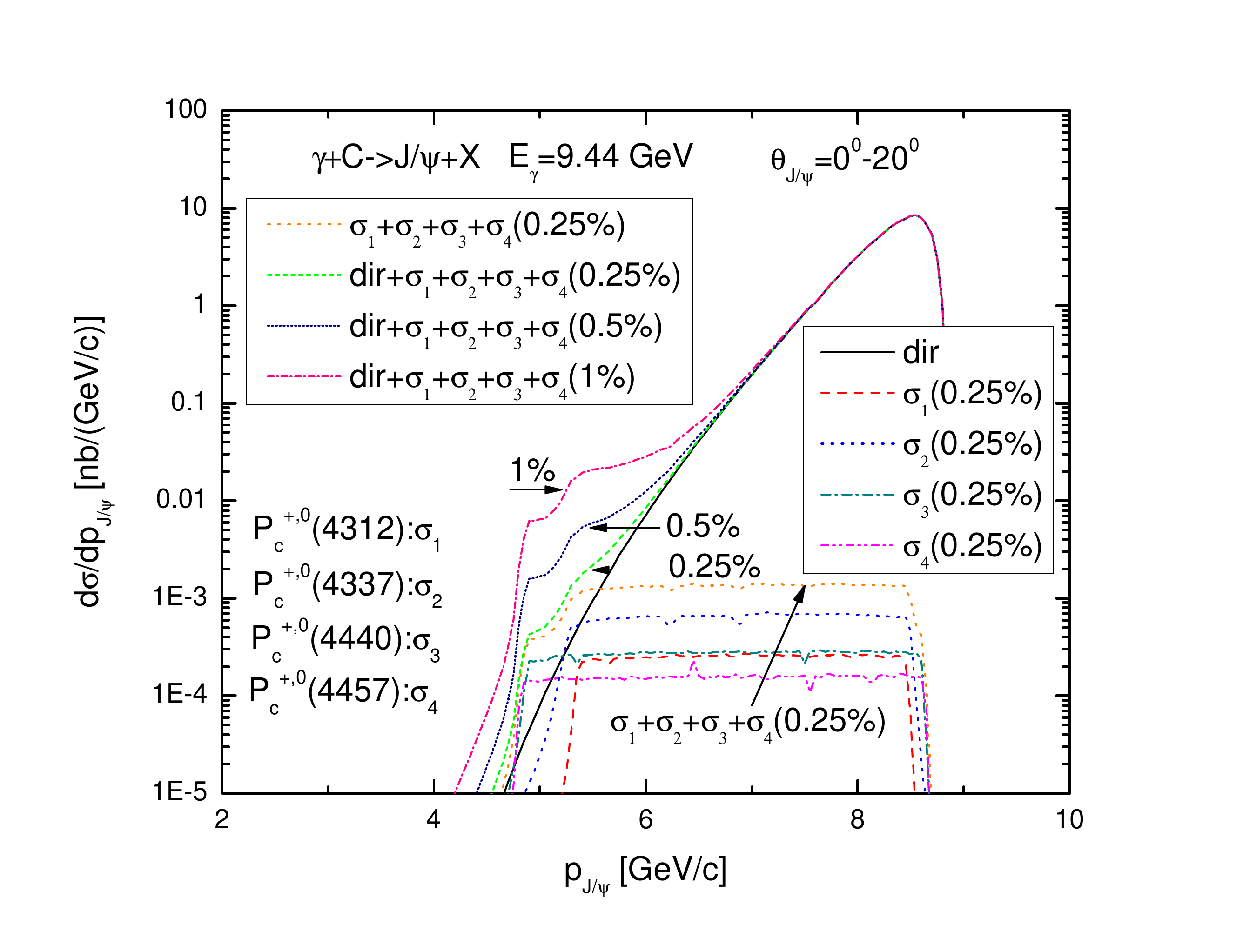}
\vspace*{-2mm} \caption{(Color online) The direct non-resonant momentum distribution
of $J/\psi$ mesons, produced in the reaction ${\gamma}{\rm ^{12}C} \to {J/\psi}X$
in the laboratory polar angular range of 0$^{\circ}$--20$^{\circ}$ and calculated in line with Eq. (24)
at initial photon resonant energy of 9.44 GeV in the laboratory system (solid curve).
The resonant momentum distributions of $J/\psi$ mesons, produced
in the two-step processes
${\gamma}p(n) \to P_{c}^+(4312)(P_{c}^0(4312)) \to {J/\psi}p(n)$,
${\gamma}p(n) \to P_{c}^+(4337)(P_{c}^0(4337)) \to {J/\psi}p(n)$,
${\gamma}p(n) \to P_{c}^+(4440)(P_{c}^0(4440)) \to {J/\psi}p(n)$ and
${\gamma}p(n) \to P_{c}^+(4457)(P_{c}^0(4457)) \to {J/\psi}p(n)$ and
calculated in line with Eq. (61) at the same incident photon energy of 9.44 GeV
assuming that the resonances $P_{c}^{+,0}(4312)$, $P_{c}^{+,0}(4337)$, $P_{c}^{+,0}(4440)$
and $P_{c}^{+,0}(4457)$ with the spin-parity assignments
$J^P=(1/2)^-$, $J^P=(1/2)^-$, $J^P=(1/2)^-$ and $J^P=(3/2)^-$, correspondingly,
all decay to the ${J/\psi}p(n)$
with branching fractions 0.25\% (respectively, red dashed, blue dotted, dark cyan dashed-doted
and magenta dashed-dotted-dotted curves) and their incoherent sum (orange dotted curve).
Incoherent sum of the direct non-resonant $J/\psi$ momentum distribution and resonant ones,
calculated supposing that the resonances
$P_{c}^{+,0}(4312)$, $P_{c}^{+,0}(4337)$, $P_{c}^{+,0}(4440)$, $P_{c}^{+,0}(4457)$
with the same spin-parity combinations all decay to the ${J/\psi}p(n)$ with branching
fractions 0.25, 0.5 and 1\% (respectively, green short-dashed, navy short-dotted and pink
short-dashed-dotted curves), all as functions of the $J/\psi$ momentum $p_{J/\psi}$ in the laboratory frame.}
\label{void}
\end{center}
\end{figure}
\begin{figure}[!h]
\begin{center}
\includegraphics[width=16.0cm]{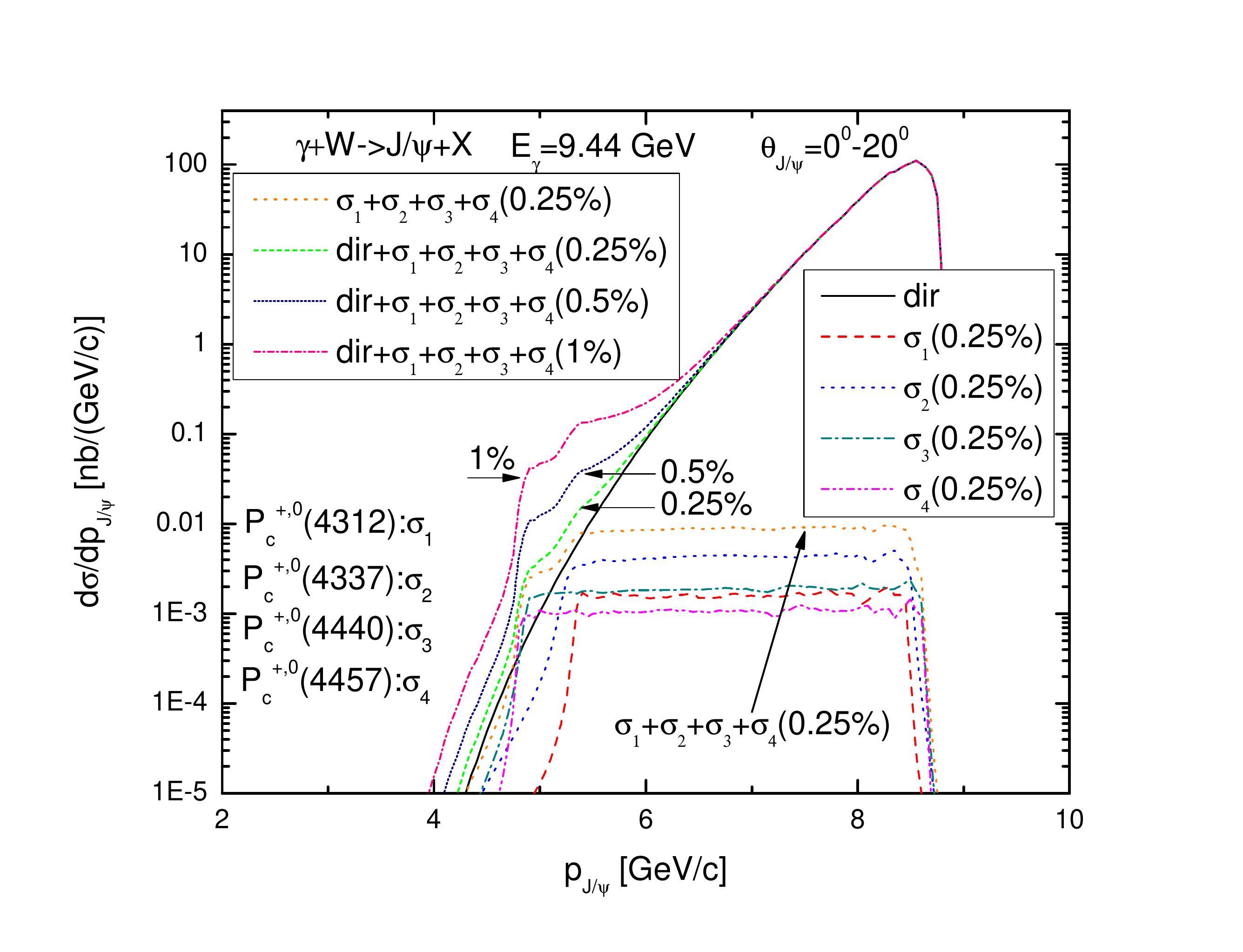}
\vspace*{-2mm} \caption{(Color online) The direct non-resonant momentum distribution
of $J/\psi$ mesons, produced in the reaction ${\gamma}{\rm ^{184}W} \to {J/\psi}X$
in the laboratory polar angular range of 0$^{\circ}$--20$^{\circ}$ and calculated in line with Eq. (24)
at initial photon resonant energy of 9.44 GeV in the laboratory system (solid curve).
The resonant momentum distributions of $J/\psi$ mesons, produced
in the two-step processes
${\gamma}p(n) \to P_{c}^+(4312)(P_{c}^0(4312)) \to {J/\psi}p(n)$,
${\gamma}p(n) \to P_{c}^+(4337)(P_{c}^0(4337)) \to {J/\psi}p(n)$,
${\gamma}p(n) \to P_{c}^+(4440)(P_{c}^0(4440)) \to {J/\psi}p(n)$ and
${\gamma}p(n) \to P_{c}^+(4457)(P_{c}^0(4457)) \to {J/\psi}p(n)$ and
calculated in line with Eq. (61) at the same incident photon energy of 9.44 GeV
assuming that the resonances $P_{c}^{+,0}(4312)$, $P_{c}^{+,0}(4337)$, $P_{c}^{+,0}(4440)$
and $P_{c}^{+,0}(4457)$ with the spin-parity assignments
$J^P=(1/2)^-$, $J^P=(1/2)^-$, $J^P=(1/2)^-$ and $J^P=(3/2)^-$, correspondingly,
all decay to the ${J/\psi}p(n)$
with branching fractions 0.25\% (respectively, red dashed, blue dotted, dark cyan dashed-doted
and magenta dashed-dotted-dotted curves) and their incoherent sum (orange dotted curve).
Incoherent sum of the direct non-resonant $J/\psi$ momentum distribution and resonant ones,
calculated supposing that the resonances
$P_{c}^{+,0}(4312)$, $P_{c}^{+,0}(4337)$, $P_{c}^{+,0}(4440)$, $P_{c}^{+,0}(4457)$
with the same spin-parity combinations all decay to the ${J/\psi}p(n)$ with branching
fractions 0.25, 0.5 and 1\% (respectively, green short-dashed, navy short-dotted and pink
short-dashed-dotted curves), all as functions of the $J/\psi$ momentum $p_{J/\psi}$ in the laboratory frame.}
\label{void}
\end{center}
\end{figure}
\begin{figure}[!h]
\begin{center}
\includegraphics[width=16.0cm]{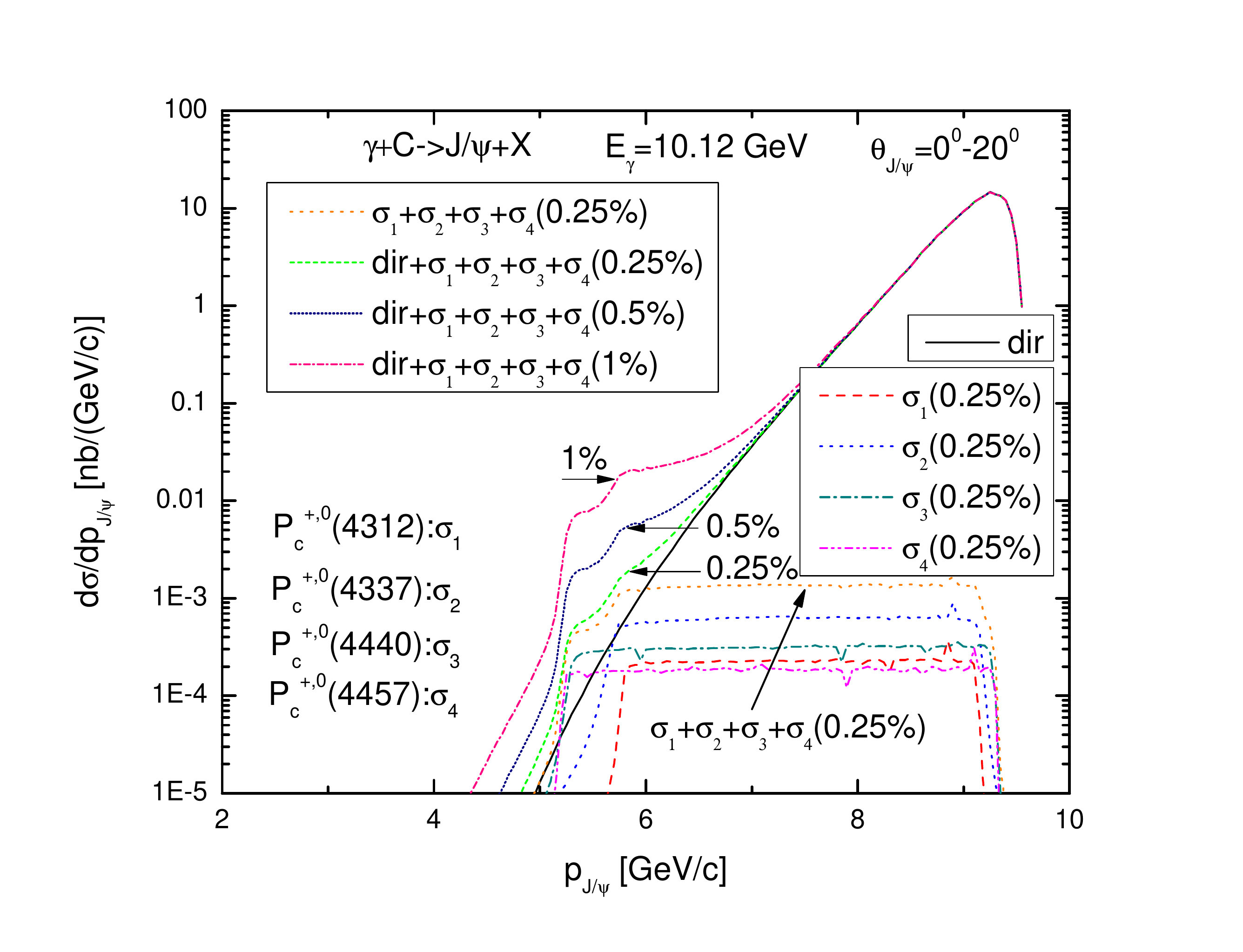}
\vspace*{-2mm} \caption{(Color online) The same as in Fig. 9, but for the initial photon energy of 10.12 GeV.}
\label{void}
\end{center}
\end{figure}
\begin{figure}[!h]
\begin{center}
\includegraphics[width=16.0cm]{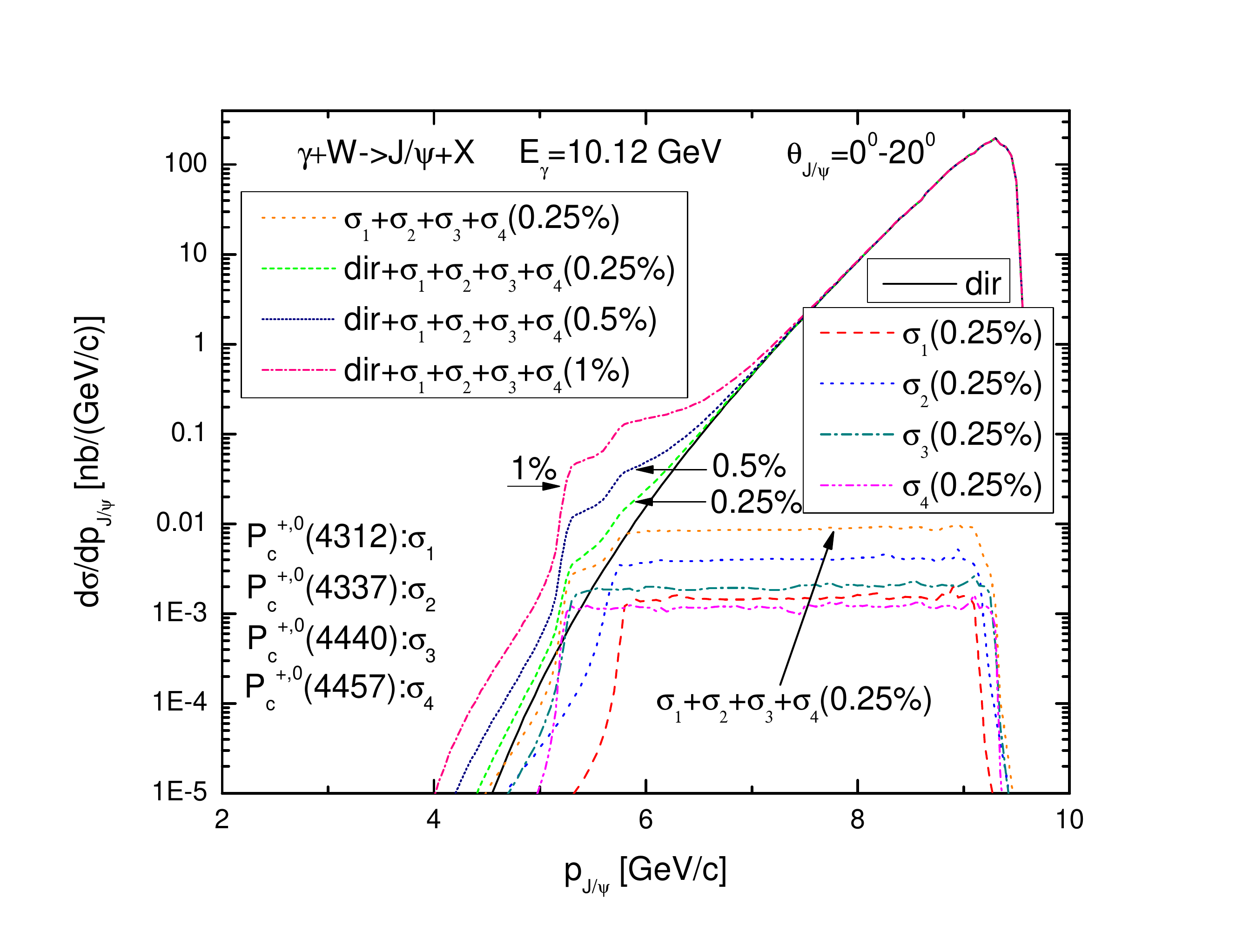}
\vspace*{-2mm} \caption{(Color online) The same as in Fig. 10, but for the initial photon energy of 10.12 GeV.}
\label{void}
\end{center}
\end{figure}

\section*{3. Results}

  The free space direct non-resonant $J/\psi$ production total cross section (23) (solid curve),
the total cross section for the resonant $J/\psi$ production in the processes (41)/(43)
determined on the basis of Eq. (52) for the considered spin-parity assignments of the hidden-charm resonances
$P_{ci}^+$ ($i=1$, 2, 3, 4) and for branching ratios $Br[P_{ci}^+ \to {J/\psi}p]=1$\% for all four
$P_{ci}^+$ states (short-dashed curve) and the combined (non-resonant plus resonant)
$J/\psi$ production total cross section (dotted curve) are presented in Fig. 4 as functions of photon energy.
It can be seen from this figure that the $P_{c}^{+}(4312)$ and $P_{c}^{+}(4337)$ as well as
$P_{c}^{+}(4440)$ and $P_{c}^{+}(4457)$ resonances exhibit itself as two narrow overlapping peaks, respectively,
at $E_{\gamma}=9.44$ and $E_{\gamma}=9.554$ GeV as well as at $E_{\gamma}=10.04$ and $E_{\gamma}=10.12$ GeV.
The strengths of these four peaks reach a value $\sim$ 0.1--0.2 nb. Whereas, the non-resonant contribution
in the resonance region is of about 1 nb. As a result, the combined total cross section of the reaction
${\gamma}p \to {J/\psi}p$ has no distinct peak structures, corresponding to the $P_{ci}^+$ states,
and it is practically not distinguished from that for the background reaction. If
$Br[P_{ci}^+ \to {J/\psi}p]=0.25$ and 0.5\%, then the resonant $J/\psi$ yield will be even more less than the
non-resonant one. This means that will be very hard to measure the $P_{ci}^+$ pentaquark states in
$J/\psi$ total photoproduction cross section on a proton target in the near-threshold energy region.
Evidently, to see their experimentally one needs to consider such observable, which is appreciably
sensitive to the $P_c^+$ signal in some region of the available phase space.
For example, the large $t$ region of the differential cross section $d\sigma/dt$
in the $J/\psi$-007 experiment [17], where the $t$-dependence of the background $J/\psi$ meson production
is suppressed while its resonant production is rather flat. This is also supported by the findings of Ref. [25],
where the photoproduction of initially claimed by the LHCb Collaboration hidden charm pentaquark states
$P_{c}^{+}(4380)$ and $P_{c}^{+}(4450)$ with the spin-parity assignments of $(3/2^-,5/2^+)$ or $(3/2^+,5/2^-)$,
respectively, on the proton target was considered by including the $t$-channel diffractive Pomeron exchanges
and the $s$-channel pentaquark productions. Here, by assuming that the pentaquark states decay into the ${J/\psi}p$
mode with fraction of 5\% was, in particular, shown that the contributions from the $P_c^+$ states calculated at
resonant c.m. energies $W=4.38$ GeV and 4.45 GeV for the two spin-parity combinations considered make the differential
cross section of the ${\gamma}p \to {J/\psi}p$ reaction strongly deviated from the diffractive one at off-forward
angles in the c.m. frame. This cross section indeed is rather flat at these angles and overestimates at them
significantly the contributions from the diffractive Pomeron exchanges. Furthermore, the predictions for the
differential cross section of the ${\gamma}p \to {J/\psi}p$ reaction, obtained in Ref. [26] within the approach
in which the Pomeron-exchange model with the parameters determined from fitting the available total cross section
data up to $W=300$ GeV is used to calculate the non-resonant amplitudes as well as the partial decay widths of
nucleon resonances with hidden charm, $N^*_{c{\bar c}}$, predicted by the considered meson-baryons ($MB$)
coupled-channel models to estimate the $N^*_{c{\bar c}} \to MB$ transition matrix elements and the vector-meson
dominance model to evaluate ${\gamma}p \to N^*_{c{\bar c}}$ as ${\gamma}p \to Vp \to N^*_{c{\bar c}}$ with
$V=\rho, \omega, J/\psi$ are adopted, demonstrate that the $N^*_{c{\bar c}}$ can be readily identified in the
near-threshold differential cross section of the ${\gamma}p \to {J/\psi}p$ process at large angles where the
contribution from Pomeron exchanges becomes insignificant. It should also be noted that an earlier prediction
of the differential cross section of this process, made in Ref. [27] at the resonant energy point $W=4.412$ GeV
by considering the non-resonant (${\gamma}p \to {J/\psi}p$) and resonant
(${\gamma}p \to N^*_{c{\bar c}}(4412) \to {J/\psi}p$) ${J/\psi}p$ photoproduction using, respectively, the two gluon
and three gluon exchange model [14] and vector-meson dominance model to generate vector mesons $\rho, \omega, J/\psi$
from photon which rescatter with the target proton to form intermediate hidden charmed nucleon resonance
$N^*_{c{\bar c}}(4412)$, shows as well that this cross section is quite weakly dependent on the c.m.s. $J/\psi$
production angle. It should be additionally pointed out that the feasibility of detecting the $P_{c}^{+}(4450)$
resonance with the spin-parity quantum numbers $J^P=3/2^-$ and $J^P=5/2^+$ in near-threshold $J/\psi$
photoproduction off protons in the CLAS12 experiment at JLab was also discussed in Ref. [23] in the framework
of a two-component model containing the directly produced resonance, diffractive background and accounting
for the experimental resolution effects. The contribution of the $P_{c}^{+}(4450)$ state, produced through
the vector-meson dominance mechanism, was parametrized using the Breit-Wigner ansatz and the non-resonant
contribution was described by the Pomeron exchanges. The fit of the available at that time data points for
differential cross section of the ${\gamma}p \to {J/\psi}p$ reaction, with $|t| \le~1.5$ GeV$^2$, covering
energy range from threshold to $E_{\gamma} \sim~120$ TeV in the lab frame, with this model showed that the upper
limits for branching ratio $Br[P_{c}^{+}(4450) \to {J/\psi}p]$ of the $P_{c}^{+}(4450)$ pentaquark range from
23\% to 30\% for $J=3/2$, depending on the experimental resolution, and from 8\% to 17\% for $J=5/2$.
These are essentially larger than those of several percent set later on by the GlueX Collaboration [4].
Finally, it is worth noting that the photoproduction of the $J/\psi$ off the proton near threshold was
studied in Ref. [28] using a novel final ${J/\psi}p$ production mechanism via the open charm $\Lambda_c^+{\bar D}^0$
and $\Lambda_c^+{\bar D}^{*0}$ intermediate states. The authors found that the existing experimental data [4]
on ${\gamma}p \to {J/\psi}p$ can be well described within the suggested mechanism. Moreover, they identified
a clear experimental signature for this mechanism: within it must be pronounced cusps at the
$\Lambda_c^+{\bar D}^0$ and $\Lambda_c^+{\bar D}^{*0}$ thresholds in the energy dependence of the total cross
section of the ${\gamma}p \to {J/\psi}p$ reaction, and found that the data [4] consistent with this feature
within their accuracy. One may hope that further measurements of the $J/\psi$ photoproduction off the proton at JLab
with higher statistics than GlueX will provide a deeper understanding of the $J/\psi$ photoproduction mechanism.

  Taking into account the aforementioned, now we consider the $J/\psi$ energy distribution
from the considered ${\gamma}p \to {J/\psi}p$ elementary reaction.
The model developed by us allows to calculate the direct non-resonant $J/\psi$ energy distribution
from this reaction, the resonant ones from the production/decay chains (41)/(43),
proceeding on the free target proton being at rest.
They were calculated according to Eqs. (39), (62), respectively, for incident
photon resonant energies of 9.44, 9.554, 10.04 and 10.12 GeV.
The resonant $J/\psi$ energy distributions were determined for the considered spin-parity assignments
of the $P_{c}^+(4312)$, $P_{c}^+(4337)$, $P_{c}^+(4440)$, $P_{c}^+(4457)$ resonances for branching fractions
$Br[P_{ci}^+ \to {J/\psi}p]=$~0.25\% for all four states.
These dependencies, together with the incoherent sum of the non-resonant $J/\psi$ energy distribution
and resonant ones,
calculated assuming that all the resonances $P_{c}^+(4312)$ and $P_{ci}^+$ ($i=1$, 2, 3, 4),
$P_{c}^+(4337)$ and $P_{ci}^+$ ($i=1$, 2, 3, 4), $P_{c}^+(4440)$ and $P_{ci}^+$ ($i=1$, 2, 3, 4),
$P_{c}^+(4457)$ and $P_{ci}^+$ ($i=1$, 2, 3, 4) decay to the ${J/\psi}p$ mode
with three adopted options for the branching ratios $Br[P_{ci}^+ \to {J/\psi}p]$,
as functions of the $J/\psi$ total energy $E_{J/\psi}$ are shown, respectively, in Figs. 5, 6, 7, 8.
It is seen from these figures that the resonant $J/\psi$ production cross sections show a flat behavior
at all allowed energies $E_{J/\psi}$. Whereas the non-resonant cross section drops fastly as $E_{J/\psi}$
decreases. At incident photon resonant energies of 9.44, 9.554, 10.04 and 10.12 GeV of interest
its strength is essentially larger than those of the resonant $J/\psi$ production cross sections,
calculated for the value of the branching ratios $Br[P_{ci}^+ \to {J/\psi}p]=$~0.25\%
for "high" allowed $J/\psi$ total energies greater than $\approx$~7.25 GeV. Whereas at "low" $J/\psi$
total energies (below 7.25 GeV) and for each considered photon energy the contribution from the resonance
with the centroid at this energy, decaying to the ${J/\psi}p$ with the branching ratio of 0.25\%,
is much larger than the non-resonant one.
Thus, for instance, in this case for the $J/\psi$ mesons with total energy of 6.5 GeV
their resonant production cross section is enhanced compared
to the non-resonant one by sizeable factors of about 2.9, 3.6, 9.5 and 22.5 at initial photon energies of
9.44, 9.554, 10.04 and 10.12 GeV, respectively. Moreover, this contribution is also substantially larger
than those, arising from the decays of another three pentaquarks to the ${J/\psi}p$ channel with the
branching ratios $Br[P_{ci}^+ \to {J/\psi}p]=$~0.25\%, at the above-mentioned "low" $J/\psi$ total energies.
As a result, at each considered photon energy the $J/\psi$ meson combined energy distribution, deriving from the
direct $J/\psi$ meson production and from the decay of the pentaquark resonance located at this energy
to the ${J/\psi}p$ mode, reveals here a clear sensitivity to the adopted  variations in the branching ratio
of this decay. Thus, for example, for the $J/\psi$ mesons with total energy of 6.5 GeV
and for the lowest considered incident photon energy of 9.44 GeV this $J/\psi$ combined distribution is enhanced
for the values of this ratio of 0.25, 0.5 and 1\% by notable factors of about 4.0, 12.5 and 46.8, respectively,
as compared to that from the directly produced $J/\psi$ mesons. And for the highest initial photon energy
of 10.12 GeV of our interest, at which the resonance $P_{c}^+(4457)$ appears as peak structure in the total
cross section of the exclusive reaction ${\gamma}p \to {J/\psi}p$, the analogous factors become much larger and
they are of about 23.5, 90.8 and 360.3, respectively. Furthermore, one can see that the above "partial"
combined energy distribution of the $J/\psi$ mesons is practically indistinguishable from their "total" combined
differential energy distribution, arising from the direct and resonant $J/\psi$ meson production via the
production/decay chains (41)/(43). This implies, on the one hand, that
the differences between the combined results, obtained by using a conservative value of the branching fractions
of the decays $P_{ci}^+ \to {J/\psi}p$ of 0.25\% and the non-resonant background, as well as between
the combined results, determined by employing the values of the branching ratios of these decays of
0.25 and 0.5\%, 0.5 and 1\%, are quite sizeable and experimentally measurable at "low" charmonium total
energies. On the other hand, at each incident
photon resonant energy considered the observation here of the specific hidden-charm LHCb pentaquark will be practically
not influenced by the presence of the another three hidden-charm pentaquark states and by the background reaction.
Since the $J/\psi$ production differential cross sections have a small absolute values $\sim$ 0.01--0.1 nb/GeV
at "low" $J/\psi$ total energies $E_{J/\psi}$, their measurement requires both high luminosities and large-acceptance detectors. Such measurement might be performed in the near future at the JLab in Hall A within the planned here high-statistics ($\sim$ 800k $J/\psi$ events in photoproduction) and high-precision E12-12-006 experiment using the SoLID detector [5, 17].

     The momentum dependencies of the absolute non-resonant, resonant and combined $J/\psi$ meson differential
cross sections, correspondingly, from the direct (1), (2), two-step (41)/(43), (42)/(44) and direct plus two-step $J/\psi$ production processes in $\gamma$$^{12}$C and $\gamma$$^{184}$W interactions, calculated on the basis of
Eqs. (24), (61) for laboratory polar angles of 0$^{\circ}$--20$^{\circ}$ and for incident photon lowest resonant energy
of 9.44 GeV, are shown, respectively, in Figs. 9 and 10. The same as in these figures, but for initial highest photon
resonant energy of 10.12, is presented in Figs. 11 and 12. The resonant momentum differential cross
sections for the production of $J/\psi$ mesons in the two-step processes
${\gamma}p \to P_{ci}^+ \to {J/\psi}p$ and ${\gamma}n \to P_{ci}^0 \to {J/\psi}n$ ($i=1$, 2, 3, 4),
proceeding on the intranuclear nucleons of carbon and tungsten target nuclei, were obtained
for three employed values of the branching ratios $Br[P_{ci}^+ \to {J/\psi}p]$ and $Br[P_{ci}^0 \to {J/\psi}n]$.
It can be seen from these figures that the total contribution to the $J/\psi$ production on both these nuclei, coming
from the intermediate $P_{ci}^+$ and $P_{ci}^0$ states decaying to the ${J/\psi}p$ and ${J/\psi}n$
modes with branching fractions of 0.25\%, shows practically flat behavior, and it is significantly larger than that from the background processes (1), (2) in the "low"-momentum regions of 4.5--5.5 GeV/c
and 4.5--6 GeV/c for considered photon beam energies of 9.44 and 10.12 GeV, respectively.
As a result, in them the combined charmonium yield is completely governed by the presence of the $P_{ci}^+$ and  $P_{ci}^0$ states in its production. Its strength is almost totally determined by the branching ratios $Br[P_{ci}^+ \to {J/\psi}p]$ and $Br[P_{ci}^0 \to {J/\psi}n]$ used in the calculations with a value, which is still large enough to be measured, as one may hope, at the CEBAF (cf. [13]), and which increases by a factor of about ten for both photon beam energies considered when going from carbon target nucleus to tungsten one
\footnote{$^)$It is interesting to note that the photoproduction of $J/\psi$-$^3$He bound state ([$^3$He]$_{J/\psi}$)
on a $^4$He target has been investigated in Ref. [29] using the impulse approximation, several
$\gamma+N \to J/\psi+N$ models based on the Pomeron-exchange and accounting for the pion-exchange mechanism
at low energies, and various $J/\psi$-nucleus potentials. The upper boundary of the predicted total cross
sections was found to be very small -- it is about 0.1--0.3 pb. The possibility of photoproduction of a six
quark-$J/\psi$ bound state ([$q^6$]$_{J/\psi}$) on the $^3$He target has been studied in Ref. [29] as
well. The upper boundary of the predicted total cross sections of $\gamma+^3{\rm He} \to [q^6]_{J/\psi}+N$
was obtained to be slightly larger than in the preceding case -- it is about 2--4 pb, depending on the model
of $\gamma+N \to J/\psi+N$ used in the calculations. These predictions may facilitate the planning of possible
measurements of [$^3$He]$_{J/\psi}$ and [$q^6$]$_{J/\psi}$ bound states at JLab.}$^)$
.
This leads to the  well separated and experimentally distinguishable
differences between all combined calculations, corresponding to the adopted options for these ratios,
for both target nuclei and for both photon energies considered.
Since the $J/\psi$ meson production differential cross sections at photon beam energy of 9.44 GeV are larger
than those at the energy of 10.12 GeV by a factor of about 20 in the above "low"-momentum regions, their measurements
on light and especially on heavy nuclear targets at photon energies in the "low"-energy resonance region will open
an opportunity to determine accurately the above branching ratios -- at least to distinguish between their realistic
options of 0.25, 0.5 and 1\%. Such measurements could also be performed in the future at the JLab in the framework
of the proposed here E12-12-006 experiment [5, 17].

  Accounting for the above considerations, we can conclude that the near-threshold $J/\psi$
energy and momentum distribution measurements in photon-induced reactions both on protons
and on nuclear targets will provide further evidence for the existence of the pentaquark
$P_{ci}^+$,  $P_{ci}^0$ resonances, and will shed light on their decay rates to the channels ${J/\psi}p$ and ${J/\psi}n$.

\section*{4. Epilogue}

In this paper we studied the near-threshold $J/\psi$ meson photoproduction from protons and nuclei
by considering incoherent direct non-resonant
(${\gamma}p \to {J/\psi}p$, ${\gamma}n \to {J/\psi}n$) and two-step resonant
(${\gamma}p \to P_{ci}^+ \to {J/\psi}p$, ${\gamma}n \to P_{ci}^0 \to {J/\psi}n$, $i=1$, 2, 3, 4;
$P_{c1}^{+,0}=P_c^{+,0}(4312)$, $P_{c2}^{+,0}=P_c^{+,0}(4337)$, $P_{c3}^{+,0}=P_c^{+,0}(4440)$,
$P_{c4}^{+,0}=P_c^{+,0}(4457)$) charmonium production processes.
We have calculated the absolute excitation functions, energy and momentum distributions
for the non-resonant, resonant and for the combined (non-resonant plus resonant) production
of $J/\psi$ mesons on protons as well as, using the nuclear spectral function approach,
on carbon and tungsten target nuclei at near-threshold incident photon energies by assuming
the spin-parity assignments of the hidden-charm resonances $P_{c}^{+,0}(4312)$,
$P_{c}^{+,0}(4337)$, $P_{c}^{+,0}(4440)$ and $P_{c}^{+,0}(4457)$ as $J^P=(1/2)^-$, $J^P=(1/2)^-$, $J^P=(1/2)^-$
and $J^P=(3/2)^-$ within three different realistic scenarios for the branching ratios
of their decays to the ${J/\psi}p$ and ${J/\psi}n$ modes (0.25, 0.5 and 1\%).
It was shown that will be very hard to measure the $P_{ci}^+$ pentaquark states through the scan of the $J/\psi$
total photoproduction cross section on a proton target in the near-threshold energy region around the resonant
photon energies of 9.44, 9.554, 10.04 and 10.12 GeV if these branching ratios $\sim$ 1\% and less.
It was also demonstrated that at these photon beam energies the $J/\psi$ energy and momentum combined distributions considered reveal distinct sensitivity to the above scenarios, respectively, at "low" $J/\psi$ total energies and momenta, which implies that they may be an important tool to provide further evidence for the existence of the pentaquark $P_{ci}^+$ and $P_{ci}^0$ resonances and to get valuable information on their decay rates to the ${J/\psi}p$
and ${J/\psi}n$ final states. The measurements of these distributions could be performed in the near future
at the JLab in Hall A within the planned here high-statistics ($\sim$ 800k $J/\psi$ events in photoproduction) and high-precision E12-12-006 experiment using the SoLID detector.

\end{document}